\documentclass[aps,onecolumn,preprint,superscriptaddress,nofootinbib,prd]{revtex4}
\usepackage{amsmath,amssymb,color,mathrsfs, graphicx,verbatim,epsfig, bbm, wasysym, multirow, url, placeins}

\def\lsim{\mathrel{\rlap{\lower4pt\hbox{\hskip1pt$\sim$}}
    \raise1pt\hbox{$<$}}}
\def\gsim{\mathrel{\rlap{\lower4pt\hbox{\hskip1pt$\sim$}}
    \raise1pt\hbox{$>$}}}

\newcommand{\be}{\begin{eqnarray}}
\newcommand{\ee}{\end{eqnarray}}

\def\addresses#1#2{\hbox to \hsize{\@tablebox{#1}\hfil\@tablebox{#2}}}
\def\@tablebox#1{\vtop{\hsize=5in \begin{flushleft} #1 \end{flushleft}}}

\def\beq{\begin{equation}}
\def\eeq{\end{equation}}
\def\bit{\begin{itemize}}
\def\eit{\end{itemize}}

\def\beqarray{\begin{eqnarray}}
\def\eeqarray{\end{eqnarray}}

\def\met{$\displaystyle{\not}E_T$}
\def\mathmet{\displaystyle{\not}E_T}
\def\PYTHIA{{\tt PYTHIA}}
\def\HERWIG{{\tt HERWIG}}
\def\MadGraph{{\tt MadGraph}}

\def\FastJet{{\tt FastJet}}

\begin{document}

\begin{titlepage}

\thispagestyle{empty}


\strut
\vspace{-1cm}

\begin{center}

\vskip 2cm

{\Large \bf Diboson-Jets and the Search for \\ Resonant $Zh$ Production}
\vskip 1.0cm
{\large Minho Son$^1$, Christian Spethmann$^2$, and Brock Tweedie$^2$}
\vskip 0.4cm
{\it $^1$Department of Physics, Yale University, New Haven, CT 06511} \\
{\it $^2$Physics Department, Boston University, Boston, MA 02215}
\vskip 2.5cm

\end{center}

\noindent
New particles at the TeV-scale may have sizeable decay rates into boosted Higgs bosons or other heavy scalars.  Here, we investigate the possibility of identifying such processes when the Higgs/scalar subsequently decays into a pair of $W$ bosons, constituting a highly distinctive ``diboson-jet.''  These can appear as a simple dilepton (plus \met) configuration, as a two-prong jet with an embedded lepton, or as a four-prong jet.  We study jet substructure methods to discriminate these objects from their dominant backgrounds.  We then demonstrate the use of these techniques in the search for a heavy spin-one $Z'$ boson, such as may arise from strong dynamics or an extended gauge sector, utilizing the decay chain $Z' \to Zh \to Z(WW^{(*)})$.  We find that modes with multiple boosted hadronic $Z$s and $W$s tend to offer the best prospects for the highest accessible masses.  For 100~fb$^{-1}$ luminosity at the 14~TeV LHC, $Z'$ decays into a standard 125~GeV Higgs can be observed with 5$\sigma$ significance for masses of 1.5--2.5~TeV for a range of models.  For a 200 GeV Higgs (requiring nonstandard couplings, such as fermiophobic), the reach may improve to up to 2.5--3.0~TeV.

\end{titlepage}

\setcounter{page}{1}



\section{Introduction}
\label{sec:intro}


The hunt for the Higgs boson is now in full swing at the LHC.  Assuming Standard Model production and decay, the available mass range has shrunk dramatically over just the past year, with a favored region emerging between about 120 and 130~GeV, and hints of signals near 125~GeV~\cite{ATLASHiggsCombined,CMSHiggsCombined,TevatronHiggsCombined}.  While this bodes well for the minimal Higgs scenario suggested by electroweak precision and flavor observables, the story is of course far from over.  Even if the Higgs is ultimately found in this range, it will immediately come under very detailed scrutiny to verify that both its production and decay rates are indeed standard.  Another possibility is that the Higgs, or one of its scalar cousins, is still hiding in the already excluded range, but with smaller rate than the Standard Model prediction.

While we continue to wait for a concrete result either way, we anticipate that the Higgs boson will be merely the first of many exciting discoveries at the LHC.  In fact, the Higgs itself will continue to play a central role in the search for new physics by serving as a signal of the production of new particles, much like gauge bosons and fermions currently serve as signals of the Higgs.  Such new physics signals are of particular interest for illuminating the full dynamics of electroweak symmetry breaking, as new particles with large couplings to the Higgs may give us clues about the Higgs's origins.  However, LHC searches for new sources of Higgs production can be nontrivial in practice, owing to the fact that Higgs decays usually involve jets.  In particular, for new particles much heavier than the Higgs, this can lead to complications in standard object reconstructions -- the jets and/or leptons from the Higgs decay can become merged due to the small angles incurred by the large transverse Lorentz boost, forming a single ``Higgs-jet.''  Therefore, it behooves us to take a closer look at how these signals might be uncovered.

Since the Higgs boson has still not been conclusively seen, and since we do not even know if its properties are strictly those predicted by the Standard Model, any phenomenological study of boosted Higgses faces an immediate two-pronged question:  What is the Higgs's mass and how does it decay?  In previous publications~\cite{Katz:2010mr, Katz:2010iq}, a subset of the present authors explored the possibility of a Standard Model Higgs near the LEP bound ($m_h \simeq 115$ GeV), identified via the decays $h\to b\bar b$ and $h\to\tau^+\tau^-$.  Utilizing the tools of jet substructure, as well as some novel modifications to tau-tagging algorithms, we found that these dijet and ditau systems could be effectively discriminated from QCD jets, even at $p_T$'s well above 1~TeV and opening angles of $O(0.1)$.  We then applied these reconstruction techniques to the search for $q\bar q \to Z' \to Zh$, where the $Z'$ is any new neutral spin-1 resonance with multi-TeV mass.  We found sensitivity in a variety of different final-state channels up to $Z'$ masses near 3~TeV, assuming a 14~TeV LHC and $O$(100~fb$^{-1}$) of luminosity.  This mass is near the electroweak precision lower bound often cited for warped extra dimension and composite Higgs models~\cite{Agashe:2003zs,Agashe:2004rs}, which had previously been estimated to be accessible for a light Higgs only after ab$^{-1}$-scale luminosity upgrades~\cite{Agashe:2007ki}.

In this paper, we extend those previous results to the much richer four-body decay modes $h \to WW^{(*)}/ZZ^{(*)} \to 4$~fermions, which we call ``diboson-jets.''  We consider in detail the two cases of a 125~GeV Higgs decaying to $WW^*$ and a 200~GeV Higgs decaying to doubly-resonant $WW$.  With the current limits from the LHC, the latter becomes viable only if this ``Higgs'' has nonstandard couplings (e.g., fermiophobic), which would already be a strong indication of additional TeV-scale physics.  It can also be a neutral scalar not directly associated with electroweak symmetry breaking, but which nonetheless decays like a heavy Higgs due to mixing~\cite{Fox:2011qc}.  For example, in the case of $Z'$ models, it may be a scalar associated with $U(1)'$ breaking.  Given that scalar $\to WW^{(*)}/ZZ^{(*)}$ decays can actually be rather generic, we expect that the techniques which we explore here will have wider applications.

The $WW^{(*)}$ diboson-jets have the usual ensemble of secondary decay modes available to systems with two $W$-bosons.  Neglecting decays with taus for simplicity, these are dileptonic ($WW^{(*)} \to l\nu l\nu$, $BR = 5$\%), semileptonic ($WW^{(*)} \to l\nu q\bar q'$, $BR = 30$\%), and fully hadronic ($WW^{(*)} \to q\bar q' q\bar q'$, $BR = 45$\%).  In the dileptonic case, care must be taken in defining lepton isolation, but otherwise their identification is straightforward~\cite{JacksonBOOST2010,CMSboostedZ}.  In the semileptonic and fully hadronic cases, we get configurations that look qualitatively similar to QCD jets, and must apply dedicated jet substructure techniques to fully reconstruct the decay kinematics.  These techniques allow us to establish that these jet-like systems in fact contain, respectively, a lepton and a subjet doublet, or a {\it quadruplet} of subjets.  In either case, the total (transverse) mass of the diboson-jet will be approximately $m_h$, and we may be able to pick out one or more on-shell $W$ subsystems.  This level of detailed reconstruction can then form the basis of dedicated semileptonic and fully hadronic diboson-jet tags, whose performance we describe in more detail below.

With these substructure techniques in hand, we demonstrate their utility for new physics searches by returning to the example of a multi-TeV $Z'$ decaying into $Zh$.  Such new resonances are ubiquitous in models where the Higgs originates as a composite particle from a new strongly-interacting sector, and these $Z'$ decays can be dominated by $Zh$ (and the $SU(2)_L$-related mode $W^+W^-$).  (See, e.g.,~\cite{Agashe:2007ki,Han:2003wu}.)  
The decay $Z' \to Zh$ also occurs with appreciable rate in simple $U(1)'$-extended gauge sectors, such as a TeV-scale hypercharge/$B-L$ admixture or bosons from $E_6$ unification (reviewed in~\cite{Langacker:2008yv}). 

We conduct a broad multi-channel survey of discovery prospects for $Z' \to Zh \to Z(WW^{(*)})$ at the 14~TeV LHC, bringing together our diboson-jet tagging techniques and well-known techniques for tagging boosted hadronic $Z$-bosons~\cite{Seymour:1993mx, Butterworth:2002tt, Butterworth:2008iy, Almeida:2008yp, Almeida:2010pa, Ellis:2009su, Krohn:2009th, Thaler:2010tr}.  We find that channels incorporating hadronic decay modes and jet substructure are the most powerful, with many different channels displaying similar performance.  The most promising amongst these at masses above 3~TeV include the ``dijet with embedded lepton'' channel $Zh \to (q\bar q)(l\nu q\bar q')$ (previously studied in~\cite{Agashe:2007ki}), as well as the ``monojet'' channel $(\nu\bar\nu)(q\bar{q}' q\bar{q}')$.  For masses below about 2~TeV, the former remains powerful, and the channels $(q\bar q)(l\nu l\nu)$, $(\nu\bar\nu)(l\nu q\bar q')$, and $(l^+l^-)(l\nu q\bar q')$ also become comparable or better.  Of course, in analogy to the ongoing Standard Model Higgs search itself, the answer to the question ``which channel is best?''~really depends on the $Z'$ mass range of interest and the available luminosity, and statistical combinations across channels can offer nontrivial benefits.  We perform a simple estimate of such a combination, and consider the discovery reach for a baseline set of $Z'$ models.  We find that a 100~fb$^{-1}$ run of the 14~TeV LHC has the potential to discover $Z' \to Zh \to Z(WW^{(*)})$ with masses of up to 1.5--2.5~TeV in the case of a 125~GeV Higgs.  For the 200~GeV Higgs, the reach potentially increases to 2.5--3.0 TeV, though the actual reach inevitably depends on the mechanism used to ``hide'' this Higgs from LHC searches.

The paper is organized as follows.  In the next section, we discuss techniques for tagging boosted hadronic and semileptonic diboson-jets.  In Section~\ref{sec:results}, we survey our studies of $Z'$ discovery prospects in a variety of different channels.  We provide some closing comments in Section~\ref{sec:conclusion}.  Finally, we include two appendices.  In Appendix~\ref{sec:details}, we give a more complete account of the technical details of our discovery estimates.  In Appendix~\ref{sec:detector}, we elaborate on our detector modeling.

\section{Tagging Diboson-Jets}
\label{sec:jetsub}

\begin{figure}[tp]
\begin{center}
\epsfxsize=0.44\textwidth\epsfbox{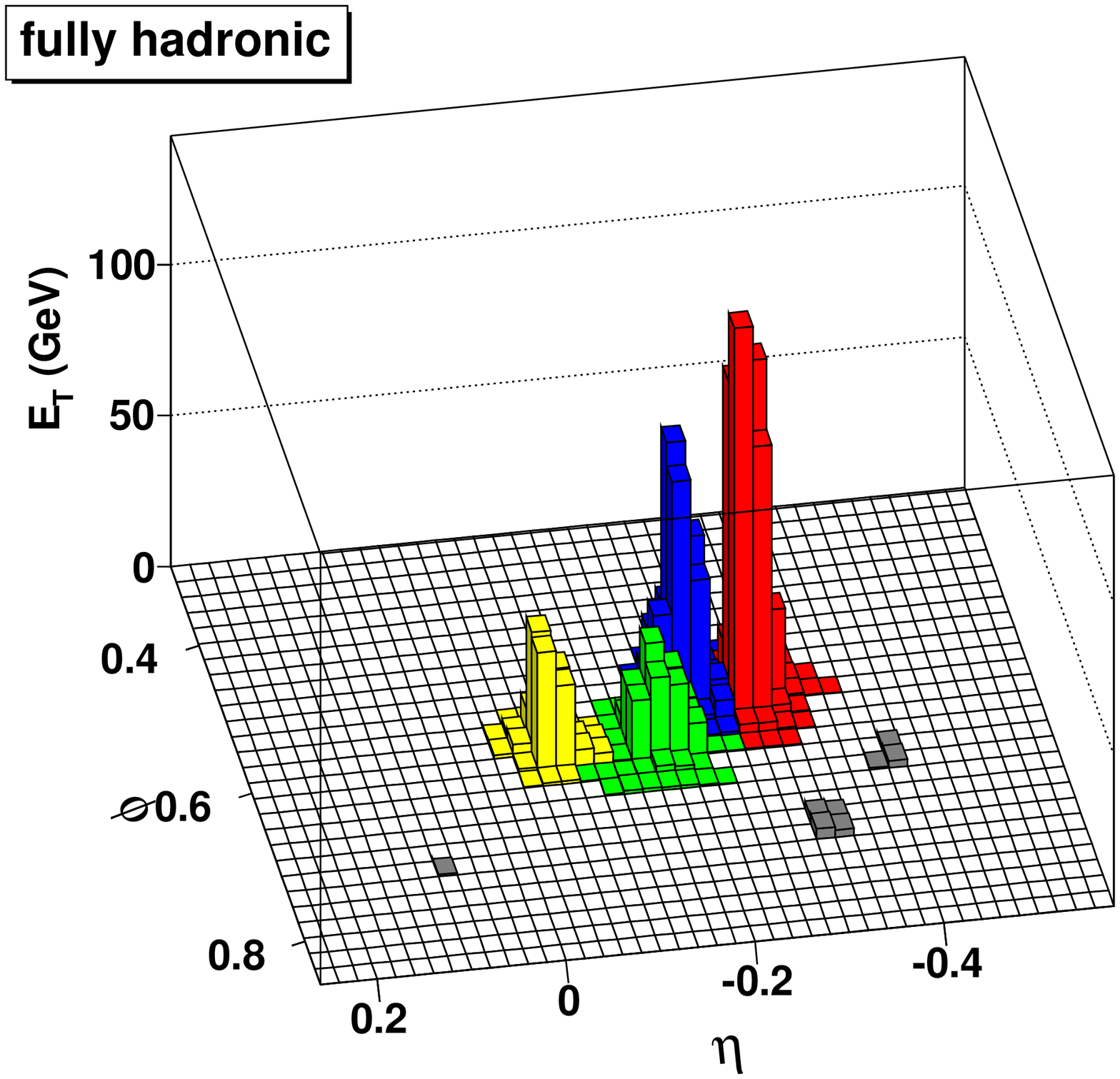}
\epsfxsize=0.44\textwidth\epsfbox{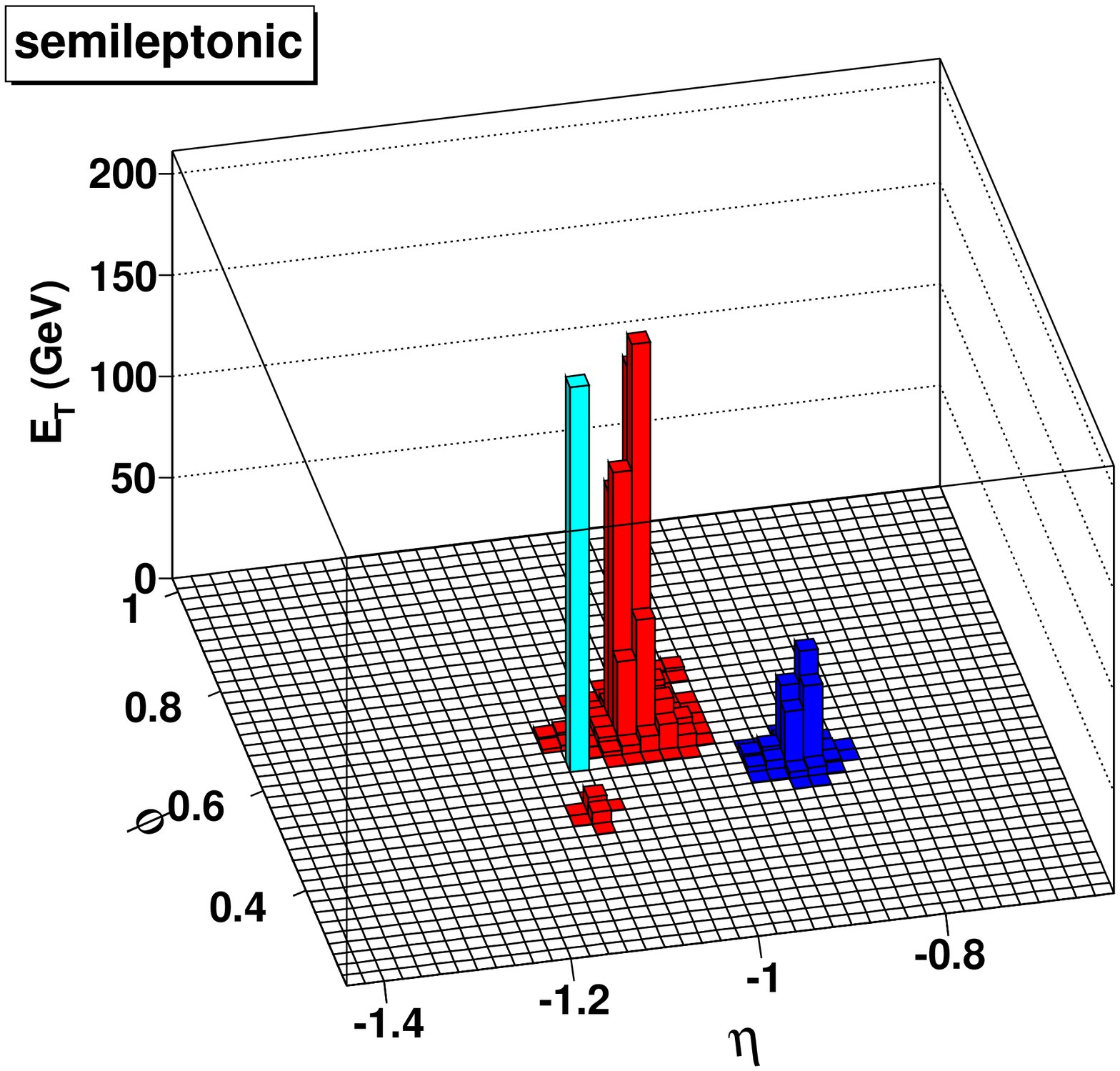}
\caption{\it Example event displays of fully hadronic (left) and semileptonic (right) diboson-jets in the $\eta$--$\phi$ plane.  These are from 125~GeV $h\to WW^{*}$ decays originating from a 3~TeV $Z'$, passed through our simple detector model and represented at ECAL-level spatial resolution.  Identified subjets are color-coded according to relative $p_T$ (in descending order:  red, blue, green, yellow).  The semileptonic example includes a non-isolated muon (cyan).  (Grey cells were thrown away by the substructure algorithm.)} 
\label{fig:legos}
\end{center}
\end{figure}

In this section, we outline some techniques for tagging highly Lorentz-boosted diboson systems, generated in the decay of a boosted Higgs or other heavy scalar.  We specialize to the case of $h\to WW^{(*)}$, which is the dominant diboson decay mode of the Higgs.  Our approach can be straightforwardly extended to $ZZ^{(*)}$, or generally any decay chain that leads to a four-body final state with any admixture of jets and leptons.

We first discuss the case of fully hadronic $WW^{(*)}$ decay, and then semileptonic decay.  We assume that the fully leptonic decay can largely be dealt with using standard lepton reconstruction, perhaps with somewhat loosened isolation criteria (as was explored for boosted leptonic $Z$ bosons in~\cite{JacksonBOOST2010,CMSboostedZ}).  Examples of fully hadronic and semileptonic diboson-jets from boosted Higgs decay can be seen in Fig.~\ref{fig:legos}.  Note that these examples, with $p_T \gsim 1$~TeV could easily fit inside of a standard-sized LHC jet ($R \geq 0.4$).\footnote{For $p_T \simeq 1$~TeV and $m_h \leq 200$~GeV, at least two fermions are merged at the $\Delta R \leq 0.4$ level in essentially every event.  The most widely separated fermions, on the other hand, can extend out to $\Delta R \sim 1$:  for the 125~GeV (200~GeV) Higgs, $\Delta R_{\rm max} \leq 1.0$ in 90\% (75\%) of events.  Lower $p_T$ and/or higher-mass diboson-jets are broader.  Below, we attempt to capture most of the decay products in a single large-radius jet that covers about half of the detector.  However, this approach ultimately becomes inefficient for $p_T \lsim 300$~GeV (500~GeV).}

To study these decay modes under semi-realistic conditions, we turn photons and hadrons into ``detector level'' objects in the form of calorimeter cells.  The finite spatial resolution of the calorimeter represents an obstacle to reliable jet substructure as we approach angular scales of $O$(0.1), exacerbated by the fact that a single particle will generally deposit energy into a cluster of nearby cells rather than staying confined to a single cell.  This tends to introduce spurious substructure and to obscure any substructure that is actually present, especially as the momentum scale exceeds a TeV.  We use the present study as an opportunity to explore methods to undo this effect.  In a typical LHC detector, the electromagnetic calorimeter (ECAL) captures an $O$(1) fraction of the total jet energy, and traces its angular distribution with $O(0.02)$ resolution.  This information can be combined with the hadronic calorimeter (HCAL), which is spatially coarser but must be included for a complete energy measurement.  We employ such a combination in the context of a highly simplified detector model.  We defer a detailed discussion to Appendix~\ref{sec:detector}, but use the results of this approach throughout the rest of the paper.

\subsection{Fully Hadronic} \label{sec:had}

The four-prong fully hadronic decay of $WW^{(*)}$ can be dealt with much in analogy to two-prong and three-prong decays such as from boosted $W$/$Z$ and $h\to b\bar b$~\cite{Seymour:1993mx, Butterworth:2002tt, Butterworth:2008iy, Almeida:2008yp, Almeida:2010pa, Ellis:2009su, Krohn:2009th, Thaler:2010tr,Almeida:2011aa} or boosted top quarks~\cite{Brooijmans:2008zza,Thaler:2008ju,Kaplan:2008ie,Almeida:2008yp,Almeida:2010pa,Plehn:2010st,Ellis:2009su,Thaler:2010tr} (see also~\cite{Abdesselam:2010pt,Altheimer:2012mn}).  Also, just as these previous techniques can be generalized to other new physics searches (e.g., R-parity violating supersymmetry~\cite{Butterworth:2009qa}, colored hadrons of a new confining sector~\cite{Bai:2011mr}), a fully hadronic four-body diboson-tagger may have other applications.  A trivial extension automatically covered by the present study is boosted $ZZ^{(*)}$, but we can also contemplate scenarios such as boosted Higgs decays to pseudoscalars $h\to aa \to 4g$ or $4b$ (also studied using substructure methods in ~\cite{Falkowski:2010hi,Bellazzini:2010uk}), or any other multi-stage decay that ends in four quarks or gluons.  For our purposes in this paper, we will simply consider the cases of a 125~GeV $h \to WW^{*}$ or 200~GeV $h \to WW$.

While the rich substructure of four-prong decays should make their discrimination from QCD-induced jets more straightforward than it is for two-prong or three-prong decays, we should also bear in mind that the energy of the parent particle has to be distributed between more objects.  This often leads to a decay with a relatively soft parton and/or a parton that is somewhat widely separated in angle from the bulk of the jet activity.  Consequently, while our first intuition may be to simply run two iterations of a two-body boosted object tagger, as is done for the Hopkins/CMS top-tagger~\cite{Kaplan:2008ie,CMStoptag}, we are often faced with situations where the first iteration distributes the partons unequally between the first level of subjets (1+3 instead of 2+2), or where the softest and/or widest-angle parton is accidentally thrown away.

To implement our diboson-tagger, we instead work with a technique that borrows some ideas from the three-body neutralino-tagger of~\cite{Butterworth:2009qa}.  (Ideas along the lines of the other multibody substructure methods cited above would also be interesting to explore.)  We first iteratively cluster the event into $R = 1.5$ quasi-hemispheric ``fat-jets'' with the Cambridge/Aachen clustering algorithm~\cite{Dokshitzer:1997in,Wobisch:1998wt}, as implemented in \FastJet~\cite{Cacciari:2005hq}.  After a fat-jet of interest is identified, we scan over its clustering history.  Working backwards, this can be viewed as a tree-like structure consisting of a sequence of $1\to2$ splittings.  Each splitting presents us with two protojet branches $j_1$ and $j_2$, with $p_T(j_1) > p_T(j_2)$ by definition.  We characterize all of the splittings with a fractional $p_T$ measure with respect to the original fat-jet, $p_T(j_2)/p_T(j_{\rm fat})$, and a mass-to-$p_T$ ratio of the softer branch, $m(j_2)/p_T(j_2)$.  If the former is smaller than a threshold $\delta_p = 0.03$, or the latter is larger than a threshold $\delta_{m/p_T} = 0.3$, we label the splitting as ``soft.''\footnote{The $m/p_T$ cut serves to eliminate would-be subjets that are really just diffuse clouds of soft radiation.  It is somewhat similar to the mass-drop criterion of~\cite{Butterworth:2008iy}, but we find that it is more effective and less redundant with the $p_T$ asymmetry cut, at least in the context of Higgs bosons with TeV-scale momenta.}  We also automatically label ``soft'' all splittings downstream of the $j_2$ branch of a soft splitting.  The remaining splittings we label ``hard.''  For these, each of the two branches carries an appreciable fraction of the jet $p_T$, the softer branch is not diffuse, and the splitting does not occur within a soft/diffuse branch.  From amongst the hard splittings, we take the three with the largest mass, $m(j_1+j_2)$.  Out of the six branches emanating from these three special splittings, there will always be exactly four branches that do not contain any of the other splittings.  We take these branches as our four {\it subjets}, discarding the rest of the particles that originally constituted the fat-jet.\footnote{It is worth pointing out that this procedure can be generalized to arbitrary numbers of desired subjets, including two or three, by simply changing the number of massive, hard splittings requested.  The two-body case essentially degenerates to~\cite{Butterworth:2008iy}, and the three-body case should overlap substantially with~\cite{Butterworth:2009qa}.  Higher numbers of subjets might be interesting to consider, for example, as a kind of alternative (and self-groomed) jet-finding algorithm that does not have a built-in minimum distance.  This is analogous to~\cite{Thaler:2010tr,Thaler:2011gf,Kim:2010uj}.}

\begin{figure}[tp]
\begin{center}
\epsfxsize=0.44\textwidth\epsfbox{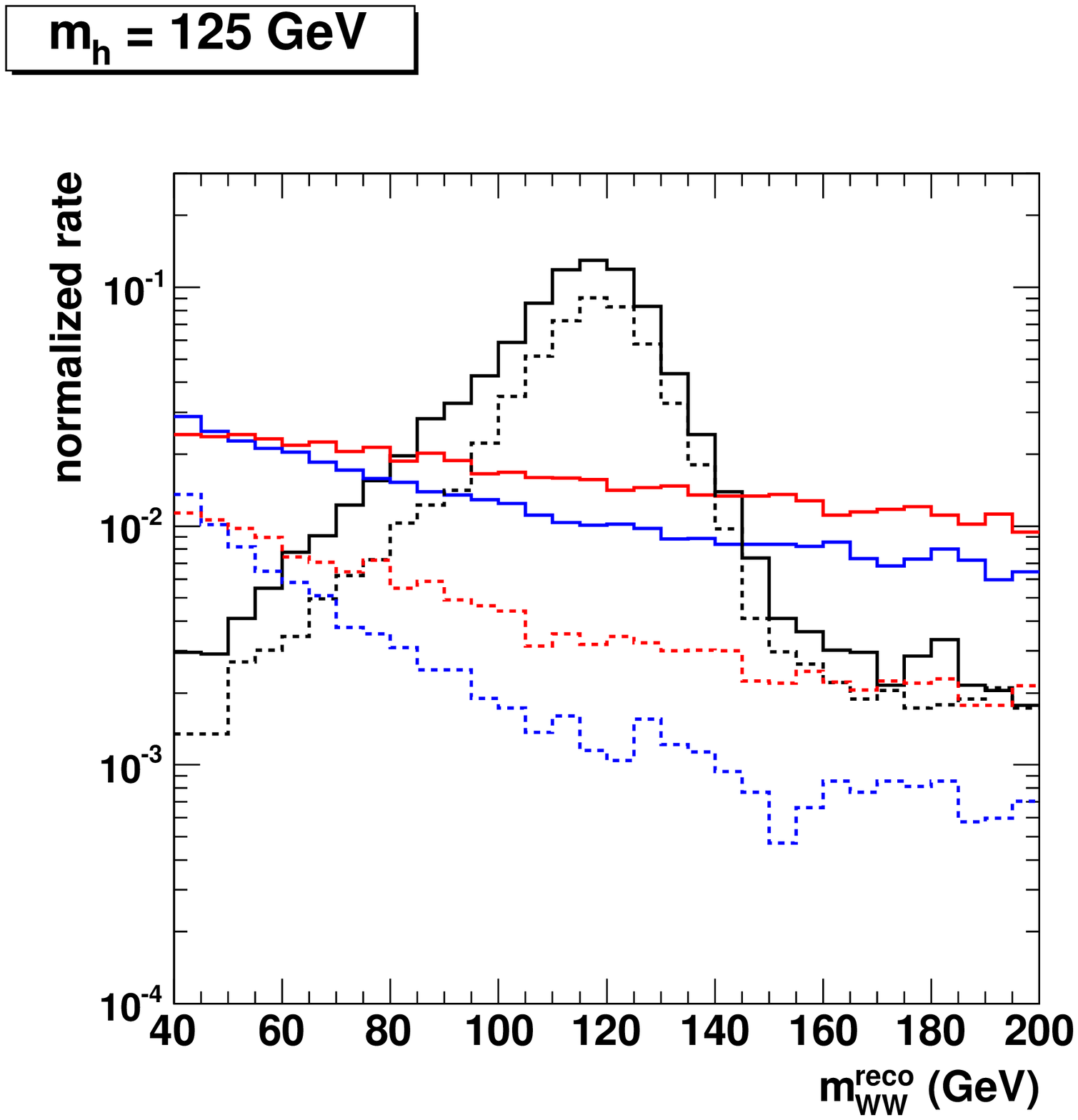}
\epsfxsize=0.44\textwidth\epsfbox{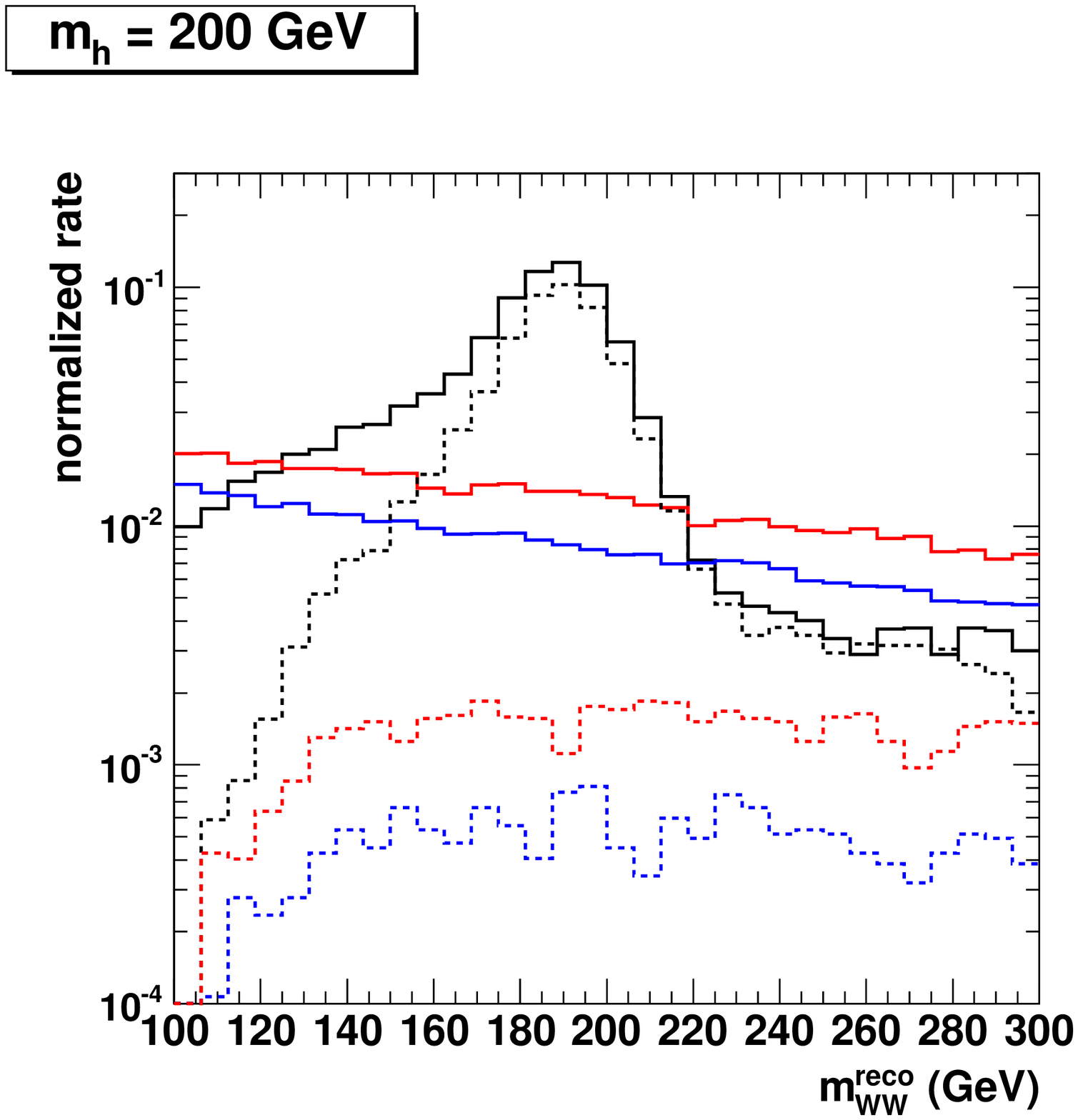}
\caption{\it Normalized distributions of reconstructed hadronic diboson mass, applying four-body substructure to 125~GeV (left) and 200~GeV (right) Higgs-jet candidates.  Signal (black) is from 2~TeV $Z'$ decay, and backgrounds of quark-jets (blue) and gluon-jets (red) are from \PYTHIA\ $Z$+jets samples of $p_T \simeq 1$~TeV.  Dashed lines indicate application of the multibody kinematic cuts discussed in the text, and are suppressed by the associated efficiencies.} 
\label{fig:mH}
\end{center}
\end{figure}

\begin{figure}[tp]
\begin{center}
\epsfxsize=0.44\textwidth\epsfbox{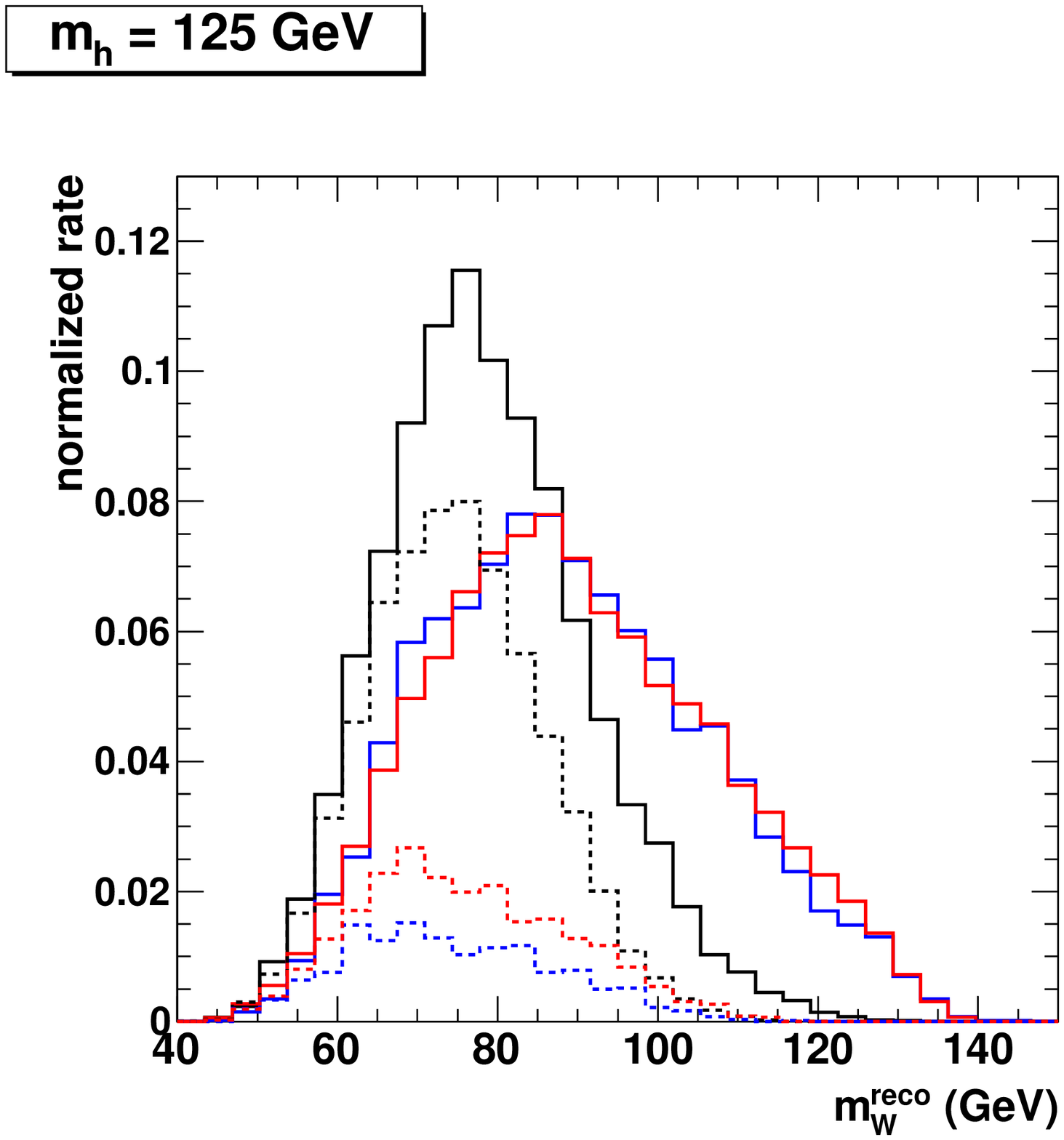}
\epsfxsize=0.44\textwidth\epsfbox{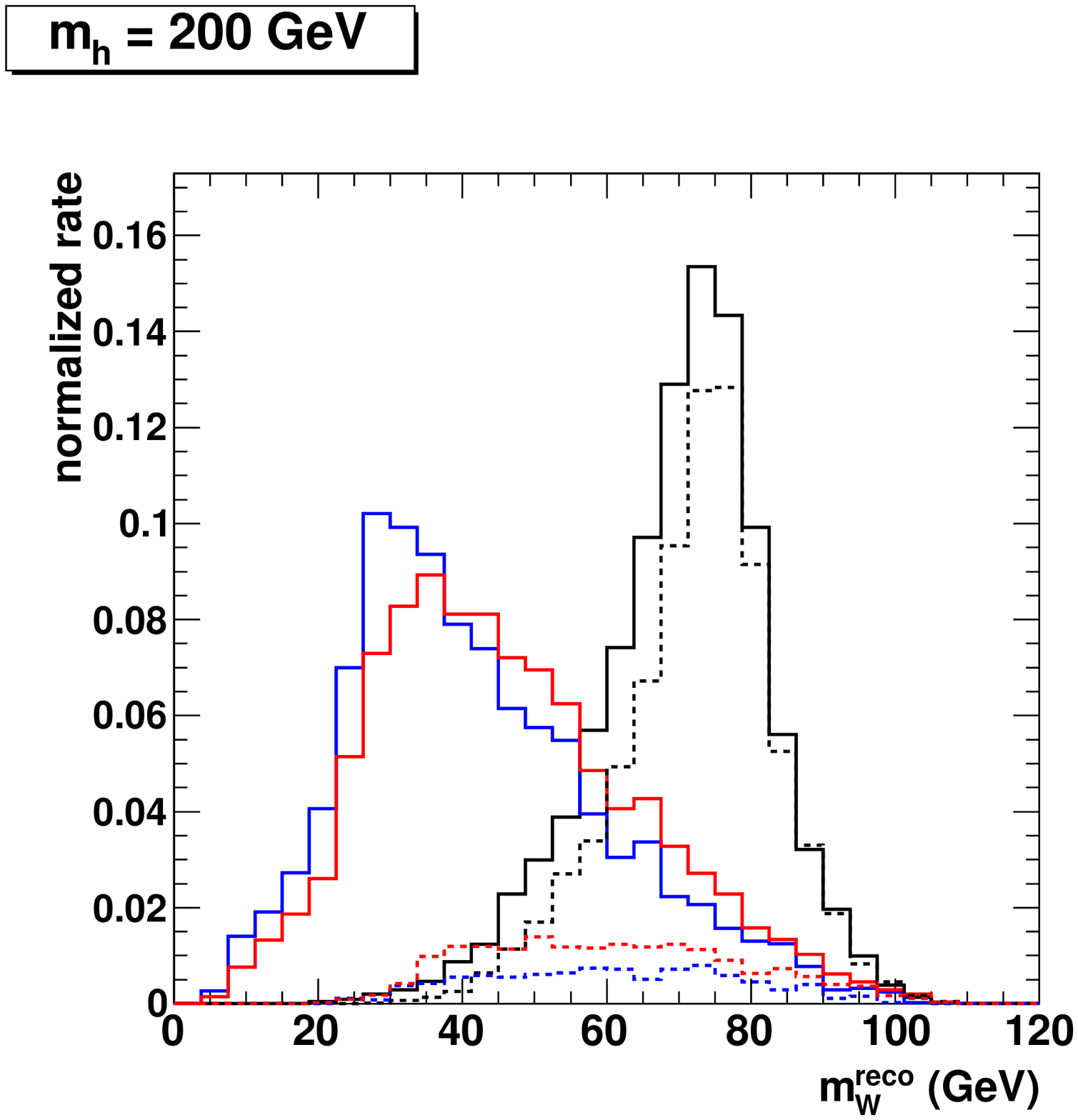}
\caption{\it Normalized distributions of reconstructed hadronic $W$ mass, applying four-body substructure to 125~GeV (left) and 200~GeV (right) Higgs-jet candidates.  Signal (black) is from 2~TeV $Z'$ decay, and backgrounds of quark-jets (blue) and gluon-jets (red) are from \PYTHIA\ $Z$+jets samples of $p_T \simeq 1$~TeV.  Events are restricted to $m_{WW}^{\rm reco} = [90,145]$ GeV or $[160,220]$ GeV, respectively.  Dashed lines indicate application of the $m_{\rm pair}^{\rm min}/m_{\rm pair}^{\rm max}$ kinematic cuts discussed in the text, and are suppressed by the associated efficiencies.}
\label{fig:mW}
\end{center}
\end{figure}

With these four subjets, we can easily reconstruct the complete diboson system mass.  This is illustrated in Fig.~\ref{fig:mH}, for fully-showered $h\to WW^{(*)}$ Monte Carlo events passed through our simple detector model.  The remaining issue is to determine whether their multibody kinematics look consistent with the partons from a $WW^{(*)}$ decay.  The kinematics will clearly be different depending on whether we are below or above the $WW$ threshold.  For the former, looking at the pair of subjets with the highest combined mass successfully reveals the on-shell $W$ peak, whereas the $W^*$ does not yield a well-localized feature.  For the latter, the doubly-on-shell decay can be revealed by considering all partitionings of the subjets into two pairs, and taking the partitioning that gives the smallest difference between the masses of the pairs.  The average of the two pair masses then typically comes out close to the $W$ mass.  Both reconstructions are illustrated in Fig.~\ref{fig:mW}.

\begin{figure}[tp]
\begin{center}
\epsfxsize=0.44\textwidth\epsfbox{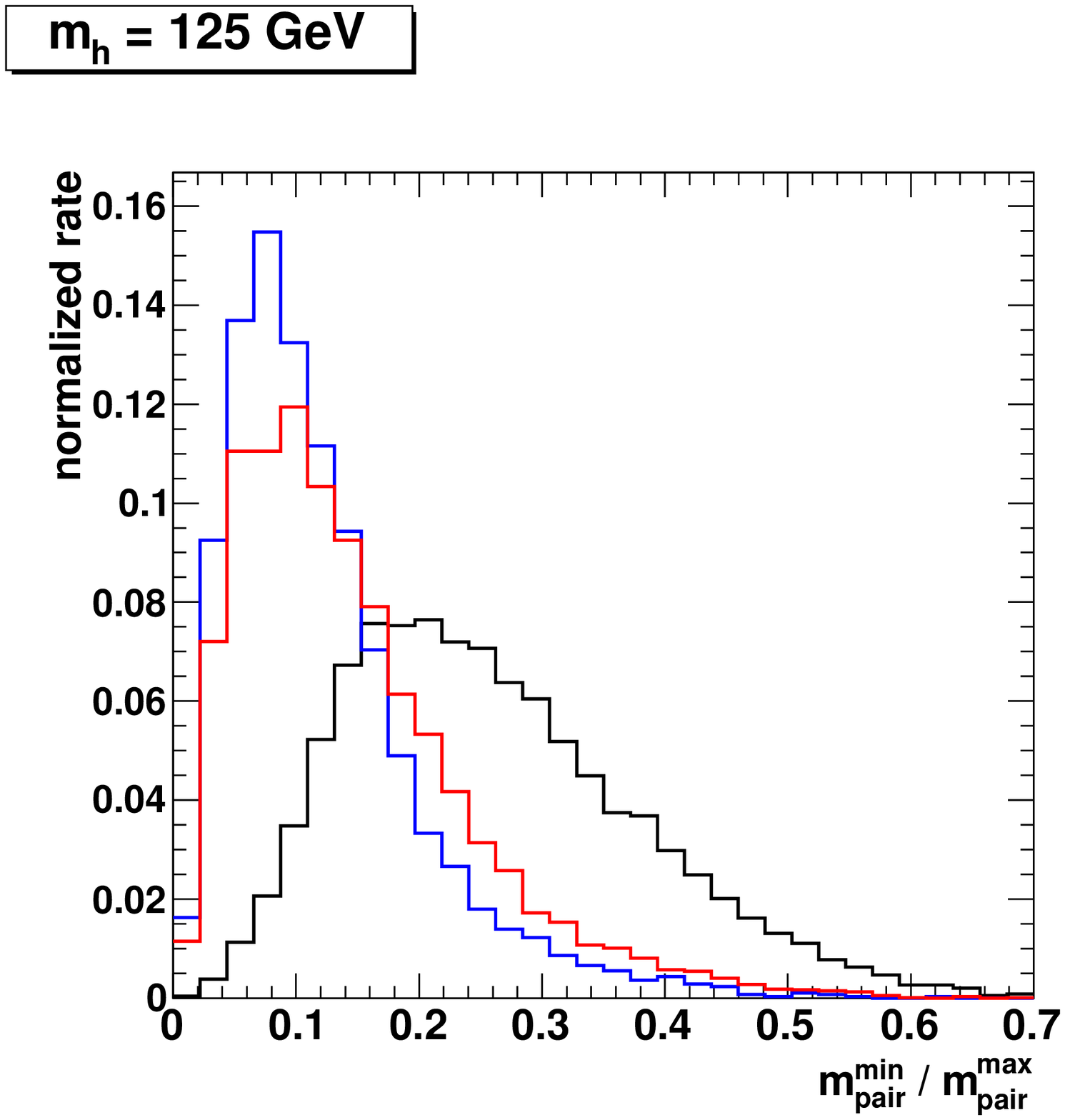}
\epsfxsize=0.44\textwidth\epsfbox{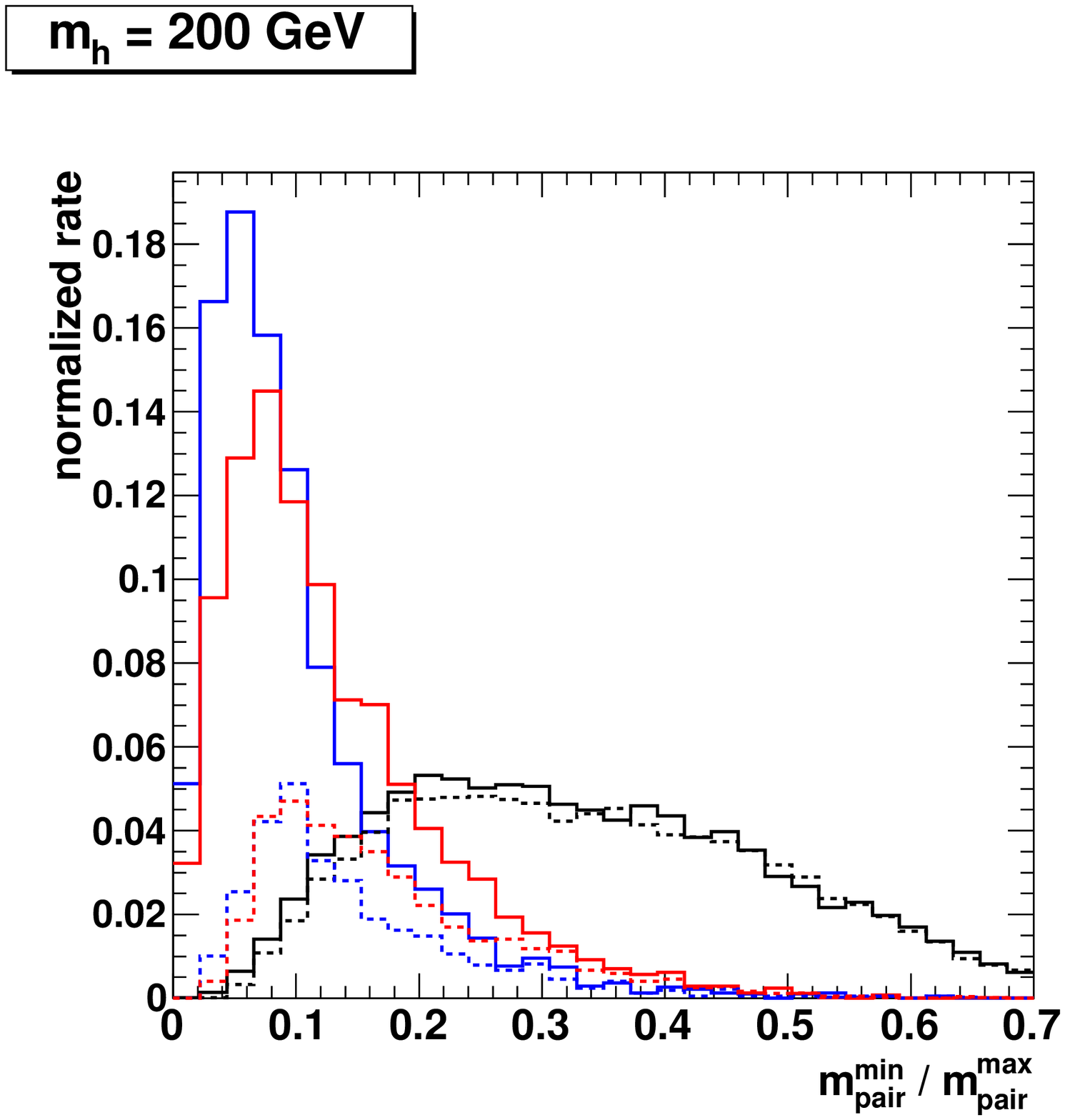}
\caption{\it Normalized distributions of the ratio between minimum subjet-pair mass ($m_{\rm pair}^{\rm min}$) and maximum subjet pair mass ($m_{\rm pair}^{\rm max}$), applying four-body substructure to 125~GeV (left) and 200~GeV (right) Higgs-jet candidates.  Signal (black) is from 2~TeV $Z'$ decay, and backgrounds of quark-jets (blue) and gluon-jets (red) are from \PYTHIA\ $Z$+jets samples of $p_T \simeq 1$~TeV.  Events are restricted to $m_h^{\rm reco} = [90,145]$ GeV or $[160,220]$, respectively.  Dashed lines indicate application of the $m_{W}^{\rm reco}$ kinematic cuts discussed in the text for the 200~GeV analysis, and are suppressed by the associated efficiencies.}
\label{fig:mMin-mMax}
\end{center}
\end{figure}

QCD jets can sometimes fake these features through parton showering.  In fact, jets that survive the above declustering procedure with mass near $m_h$ almost automatically contain subjet pairs with mass near $m_W$.  By studying fully showered jets in \PYTHIA\ and \HERWIG, we have found that a simple additional discriminating variable is the dimensionless ratio $m_{\rm pair}^{\rm min}/m_{\rm pair}^{\rm max}$, taken between the subjet pairs with the minimum and maximum invariant mass.  (The pairs need not be exclusive with respect to each other.)  The distributions can be seen in Fig.~\ref{fig:mMin-mMax}.  For our 125~GeV working point, we found that a minimum cut of $0.2$ is quite effective.  Such a cut appears to subsume any cuts on the $W$ mass, while being strictly more powerful.  The pairwise mass ratio is also quite effective for our 200~GeV working point, and is complementary to a cut demanding that the average pair mass is near $m_W$.\footnote{Of course, the full four-body phase space of the decay at either mass point is quite rich, and in principle a more aggressive multivariate approach could be even more powerful.  It might also be possible to fold in detailed information characterizing the distributions of particles in and around the subjets (see, e.g., \cite{Gallicchio:2010sw,Gallicchio:2010dq,Cui:2010km,Han:2011ab,Hook:2011cq}).}

The full tagger is then built on the following kinematic cuts.  For the 125~GeV Higgs, we demand that the four-subjet mass lies in the window $[90,145]$ GeV, and that $m_{\rm pair}^{\rm min}/m_{\rm pair}^{\rm max} > 0.2$.  For the 200~GeV Higgs, we require a window of $[160,220]$ GeV, $m_{\rm pair}^{\rm min}/m_{\rm pair}^{\rm max} > 0.2$, and $m_W^{\rm reco} > 50$~GeV, according to the above prescription.

\begin{table}[tp]
 \centering
\begin{tabular}{ c|c c c }
          &  $\;\; p_T \simeq 500$ GeV $\;\;$ & $\;\; p_T \simeq 1000$ GeV $\;\;$ & $\;\; p_T \simeq 1500$ GeV $\;\;$  \\ \hline
$\; \; h$ (125 GeV) $\; \;$  &  0.44 \      &  0.49 \ \ \ \  &  0.51 \ \ \ \  \\ \hline
quark $\to h$\  &   0.015 \, (0.018)  &   0.018 \, (0.024) &   0.027 \, (0.036) \\
gluon $\to h$\  &   0.039 \, (0.055)  &   0.040 \, (0.056) &   0.043 \, (0.054) \\
\end{tabular} 
\caption{\it Tag rates for 125 GeV boosted Higgs bosons decaying to hadronic $WW^*$, and mistag rates for \PYTHIA\ $($\HERWIG$)$ QCD jets.}
\label{tab:eff_125}
\end{table}

\begin{table}[tp]
 \centering
\begin{tabular}{ c|c c c }
          &  $\;\; p_T \simeq 500$ GeV $\;\;$ & $\;\; p_T \simeq 1000$ GeV $\;\;$ & $\;\; p_T \simeq 1500$ GeV $\;\;$  \\ \hline
$\; \; h$ (200 GeV) $\; \;$  &  0.39 \       &  0.50 \ \ \ \    &  0.51 \ \ \ \   \\ \hline
quark $\to h$\  &   $\; \; 4.9 \! \times \! 10^{-3}$ ($7.2 \! \times \! 10^{-3}$) $\; \;$ 
                &  $\; \; 5.6 \! \times \! 10^{-3}$  ($8.4 \! \times \! 10^{-3}$) $\; \;$ 
                &  $\; \; 6.6 \! \times \! 10^{-3}$  (0.010) $\; \;$ \\
gluon $\to h$\  &   0.013 \, (0.020)  &   0.016 \, (0.026)  &   0.016 \, (0.027) \\
\end{tabular} 
\caption{\it Tag rates for 200 GeV boosted Higgs bosons decaying to hadronic $WW$, and mistag rates for \PYTHIA\ $($\HERWIG$)$ QCD jets.}
\label{tab:eff_200}
\end{table}

We indicate the tag rates for boosted Higgses and mistag rates for quark- and gluon-jets in Tables~\ref{tab:eff_125} and~\ref{tab:eff_200}.  (The fractional statistical errors on these numbers are percent-scale.)  Signal samples are from {\tt MadGraph v4.5.1}~\cite{Alwall:2007st} interfaced with \PYTHIA, and background samples are from \PYTHIA\ {\tt v6.4.15}~\cite{pythiamanual} (virtuality-ordered) and \HERWIG\ {\tt 6.520}~\cite{Corcella:2000bw} (interfaced with {\tt JIMMY}~\cite{Butterworth:1996zw}).  

The tagger has a typical signal acceptance of about 50\%, and mistag rates that range from $0.5$\% to 5\%.  Generally speaking, gluon-jets have 2--3 times higher mistag rates than quark-jets, and \HERWIG\ (angle-ordered) jets can have roughly 50\% higher mistag rates than \PYTHIA\ (virtuality-ordered) jets.  Mistag rates for the 125~GeV $h\to WW^{*}$ are about 2--3 times bigger than for the 200~GeV $h\to WW$.  Removing the detector model and reverting to particle level decreases mistag rates by about a factor of 2, mainly due to better resolution on $m_{\rm pair}^{\rm min}$.  Tag and mistag rates are both fairly stable versus $p_T$.

\subsection{Semileptonic} \label{sec:semilep}

When a boosted $WW^{(*)}$ pair decays semileptonically, we in principle get a much cleaner tag, assuming that we can reliably identify the lepton in the presence of nearby hadronic activity.  For 1~(3)~TeV $Z'$ decay into a $Z$ and a 125~GeV Higgs, the lepton is more than $\Delta R = 0.4$ away from the accompanying jet activity in only 40\%~(5\%) of events, so normal isolation-based lepton identification will often fail.  The issue of lepton identification inside of a jet has been dealt with before in the context of semileptonic top-jets~\cite{Agashe:2006hk,Thaler:2008ju,Rehermann:2010vq,ATLASsemilep2009,ATLASsemilep2010}.  There it has been pointed out that the high mass scale of the decay can be exploited to efficiently separate out signal leptons from heavy flavor decays or fakes, even for very small $\Delta R$ scales.  The separation appears to be good enough that, to first approximation, it is valid to ignore these sources of background.  In particular,~\cite{Rehermann:2010vq,ATLASsemilep2010} showed that for dijet-like configurations with a lepton near one of the jet axes, electroweak bremsstrahlung of a leptonic $W$ was a much bigger worry than heavy flavor decays. 

The main technique which we utilize for lepton identification is {\it mini-isolation}, introduced in~\cite{Rehermann:2010vq}.  Real or fake leptons produced inside of very energetic jets are almost always accompanied by nearby showering and decay products.  For the specific case of leptons from heavy flavor decay, the relevant $\Delta R$ scale is set by the parent quark's mass-to-$p_T$ ratio.  While this can become quite small --- much smaller than a standard lepton isolation cone --- an isolation variable built using a miniature cone tailored to this $\Delta R$ and tallying only tracks should be quite robust for vetoing heavy flavor up to almost arbitrarily high energy.\footnote{Because this form of isolation is based solely on tracks, it may also be less sensitive to pileup than calorimeter-based isolation.}  In contrast, leptons produced from boosted top or Higgs decays have a much larger intrinsic $\Delta R$ scale for the same $p_T$, and are therefore fairly clear of extra radiation at the scale of the miniature cone associated with heavy flavor.

More explicitly, mini-isolation utilizes a cone that scales inversely with the lepton $p_T$: $R_{iso} = (15\;{\rm GeV})/p_T(l)$.  The $p_T$ of all tracks inside of this cone is scalar-summed, and the sum not allowed to exceed 10\% of the lepton $p_T$.  While the original version of mini-isolation was applied only to muons, subsequent experimental simulations have applied the same technique to electrons and found simliarly good performance~\cite{ATLASsemilep2010}.  Still, electron identification requires much more rigorous quality criteria, for example to discriminate against charged pions that deposit a large fraction of their energy in the ECAL.  We cannot properly reproduce these quality criteria in our own highly simplified analysis, but we do restrict the size of the mini-isolation cone to be larger than $0.1$ for electrons.  This in principle gives enough space to properly identify the electron track and its associated tracker/ECAL shower.\footnote{We do not attempt to impose ``isolation'' within the ECAL, which is also necessary for clean electron ID.  However, tracker isolation is in any case highly correlated with electromagnetic isolation.}

Besides the embedded lepton, semileptonic diboson-jets also have an additional layer of substructure that we can exploit, in the form of the accompanying subjet-doublet which we reveal through the ``mass drop'' algorithm introduced by Butterworth, Davison, Rubin, and Salam (BDRS)~\cite{Butterworth:2008iy}.  For partially off-shell decays, the subjets may have a mass near $m_W$ or may occupy the continuum not far below $m_W$.  For double-on-shell decays, the subjets will always have a mass near $m_W$. In either case, both the cluster transverse mass of the $l\nu q\bar q'$ system and the total visible mass of the $lq\bar q'$ system will be close to $m_h$.  This potentially gives us a lot of kinematic discrimination power against backgrounds. 

\begin{figure}[tp]
\begin{center}
\epsfxsize=0.44\textwidth\epsfbox{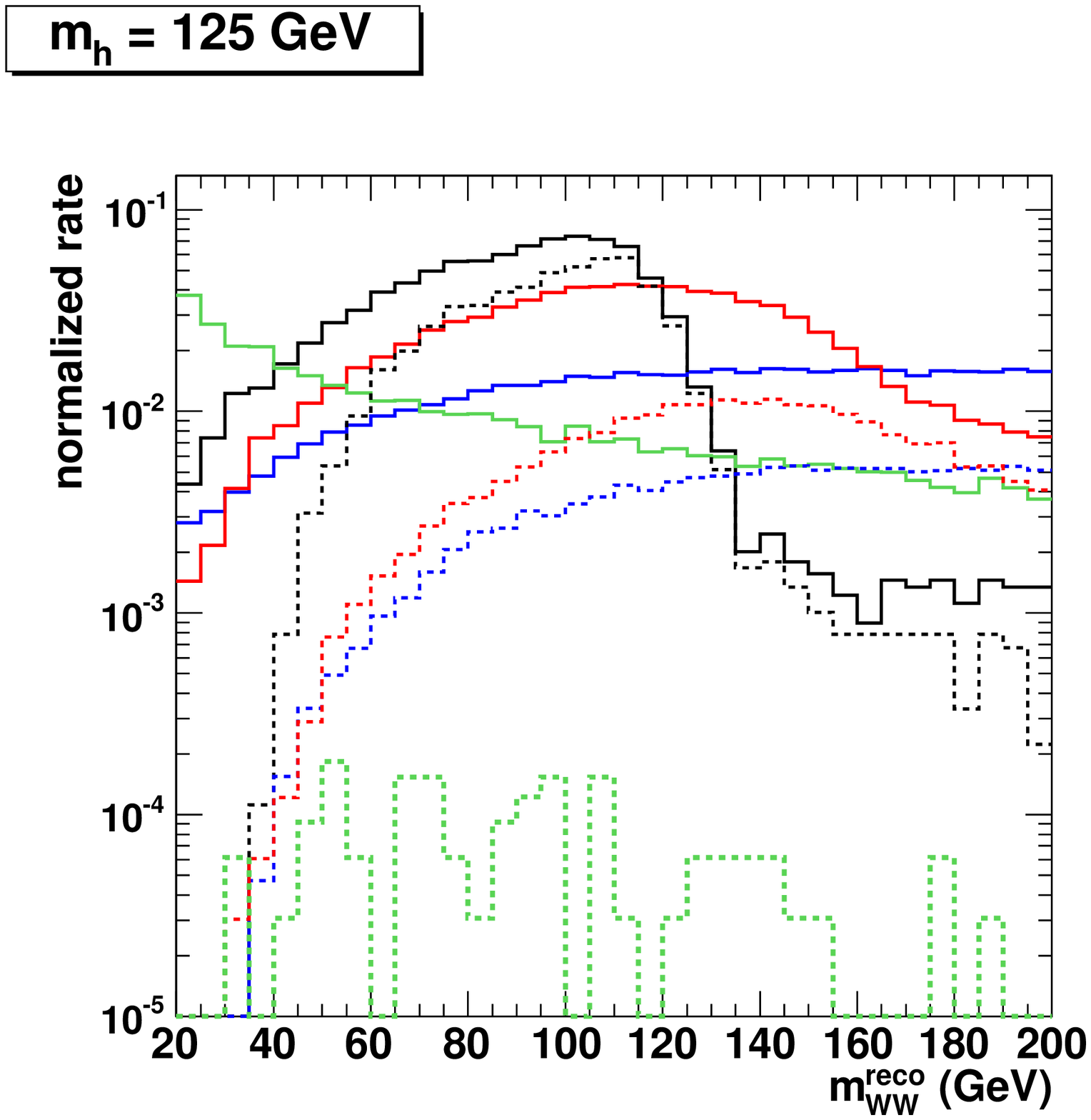}
\epsfxsize=0.44\textwidth\epsfbox{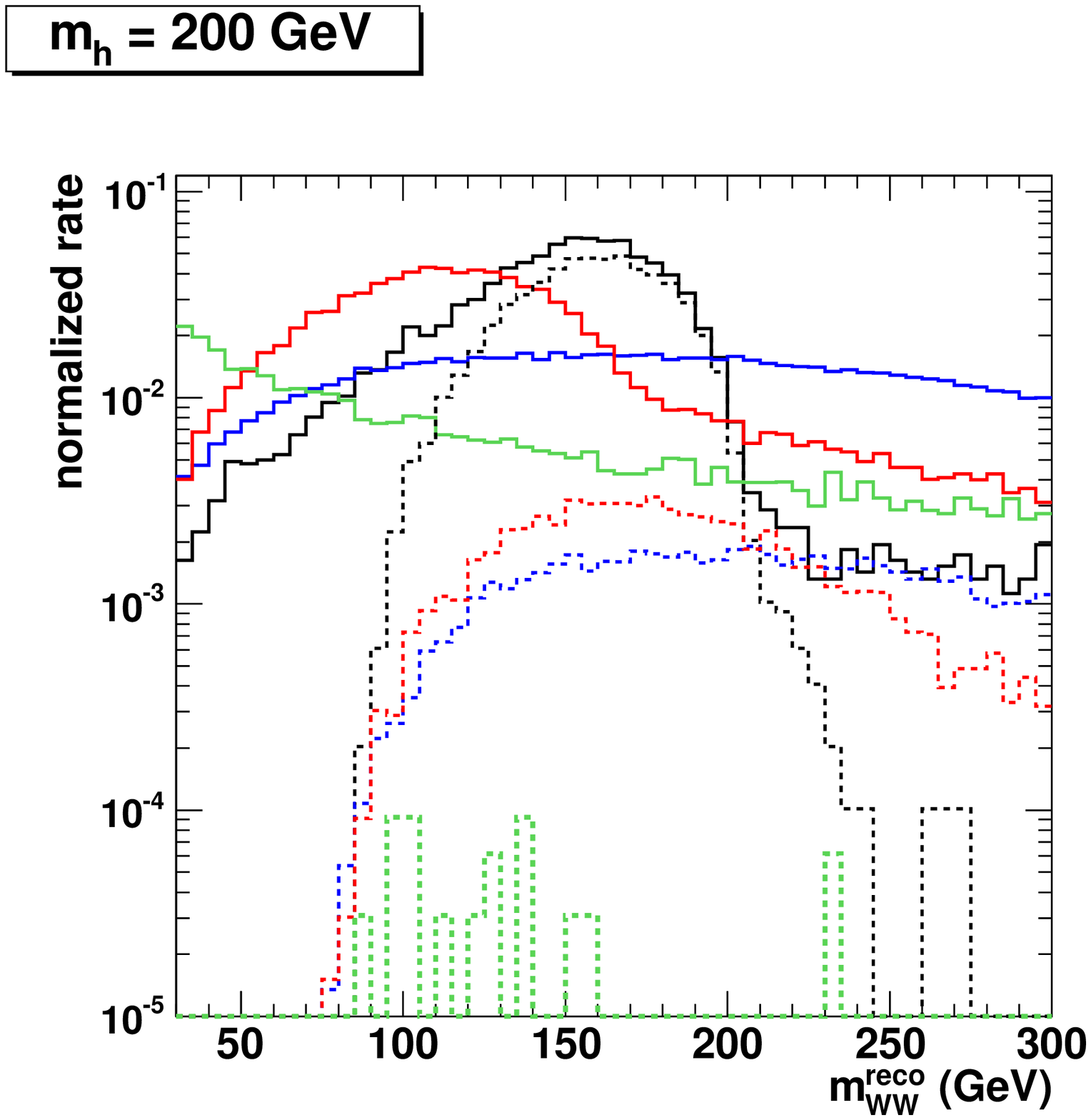}
\caption{\it Normalized distributions of reconstructed diboson-jet mass distributions, applying semileptonic substructure to 125~GeV (left) and 200~GeV (right) Higgs-jet candidates.  Signal (black) is from 2~TeV $Z'$ decay, and backgrounds of $W$-strahlung (blue), semileptonic top-jets (red), and QCD jets with embedded leptons (green) are from {\tt MadGraph5} or standalone \PYTHIA\ dijet samples of $p_T \simeq 1$~TeV.  Dashed lines indicate application of the mini-isolation and substructure kinematic cuts discussed in the text, and are suppressed by the associated efficiencies.}
\label{fig:mTcl}
\end{center}
\end{figure}

More specifically, in cases where the diboson-jet recoils against visible activity (e.g., $Z \to l^+l^-$ or $q\bar q$), the missing energy vector is a faithful tracer of the neutrino's $\vec{p}_T$, allowing a complete reconstruction in principle.  Here, we take a somewhat more minimalistic approach by simply setting the rapidity of the neutrino to match that of the four-vector sum of the lepton and the two subjets.  Given the possible uncertainties in determining the exact pointing of the \met\ vector, and the high sensitivity of our reconstruction to even $O(0.1)$ errors in angle, we also conservatively align the neutrino with the lepton+subjets in $\phi$.  We can then take the invariant mass of this system as a proxy for the diboson candidate mass.  The result of this procedure is illustrated in Fig.~\ref{fig:mTcl}, which displays a clear peak near $m_h$.  In cases where the diboson-jet recoils against a system with additional neutrinos (such as $Z \to \nu\bar\nu$), the \met\ no longer gives us a clear indication of the $p_T$ of the neutrino in the $WW^{(*)}$ decay, and we must resort instead to the visible mass.  While this is also a good mass estimator, the distribution is broader, and there is 2--3 times more background under the signal peak from background events which would have otherwise had too-high $m^{\rm reco}_{WW}$.

By combining lepton mini-isolation, two-body hadronic substructure, and lepton/subjet kinematics, we can form our semileptonic diboson tagger.  Leptons are matched to the closest fat-jet in $\Delta R$.  (As in the fully hadronic case, fat-jets are reconstructed with $R = 1.5$.)  For the case of a 125~GeV Higgs, we require the subjet-pair mass to lie in the window $m^{\rm reco}_W = [20,100]$~GeV and that the reconstructed diboson mass should be less than 130~GeV.  (The same diboson mass criterion is applied regardless of whether we can use the neutrino or not.)  To further reduce backgrounds from QCD, we also require $\Delta R(l,j_W) > (30\;{\rm GeV})/p_T(lj_W j_W)$, where $j_W$ refers to either hadronic $W^{(*)}$-subjet.  For the 200~GeV Higgs, the analogous cuts are $m^{\rm reco}_W = [60, 100]$~GeV and $m^{\rm reco}_{WW} < 200$~GeV, as well as $\Delta R(l,j_W) > (50\;{\rm GeV})/p_T(lj_W j_W)$.

In Tables~\ref{tab:eff_125h_semi} and~\ref{tab:eff_200h_semi}, we show the tag rates for semileptonic Higgs-jets and QCD jets with an embedded hard lepton.  We consider three mechanisms for lepton production inside of jets:  virtual $W$ emission in hadron decays (dominated by radiatively-produced $b$ and $c$ quarks), real $W$ bremsstrahlung from left-handed quarks, and semileptonic decays of boosted top quarks.  We model the first using \PYTHIA\ dijets, including radiative and prompt heavy flavor contributions, as well as decays-in-flight of light mesons.  (These samples contain a central lepton with $p_T > 25$~GeV in approximately 5\% of events with TeV-scale jets.)  Real $W$-strahlung samples are generated from $(W\to l\nu)jj$ simulations in {\tt MadGraph5} {\tt v1.3.30}~\cite{Alwall:2011uj}.  Semileptonic top samples come from 6-body $t\bar t \to (l\nu b)(q\bar q'b)$ simulations, also in {\tt MadGraph5}.

\begin{table}[tp]
 \centering
\begin{tabular}{ l|c c c }
          &  $\; \; p_T \simeq 500$ GeV $\; \;$ & $\;\; p_T \simeq 1000$ GeV $\;\;$ & $\;\; p_T \simeq 1500$ GeV $\;\;$  \\ \hline
   $h$ (125 GeV)                 &  0.48 \, (0.84)       &  0.45 \, (0.78)   &  0.43 \, (0.76)  \\  \hline
   $jl \to h$ (virtual $W$) $\; \;$  &  $\; \; 1.0 \! \times \! 10^{-3}$  (0.012) $\; \;$ 
                                     &  $\; \; 2.6 \! \times \! 10^{-4}$ ($7.3 \! \times \! 10^{-3}$) $\; \;$  
                                     &  $\; \; 7.1 \! \times \! 10^{-5}$  ($4.8 \! \times \! 10^{-3}$) $\; \;$ \\
   $jl \to h$ (real $W$)          &  0.044 \, (0.96) &\   0.034 \, (0.95)   &\   0.024 \, (0.94) \\
   $t_{\rm semilep} \to h$                     &  0.084 \, (0.89) &\   0.07 \, (0.85)   &\   0.069 \, (0.83)
\end{tabular} 
\caption{\it Tag rates for 125 GeV boosted Higgs bosons decaying to semileptonic $WW^*$ and mistag rates for QCD jets with an embedded lepton.  (The numbers in parentheses are tag rates for mini-isolation alone.)}
\label{tab:eff_125h_semi}
\end{table}

\begin{table}[tp]
 \centering
\begin{tabular}{ l|c c c }
          &  $\;\; p_T \simeq 500$ GeV $\;\;$ & $\;\; p_T \simeq 1000$ GeV $\;\;$ & $\;\; p_T \simeq 1500$ GeV $\;\;$  \\ \hline
   $h$ (200 GeV)                &  0.49 \, (0.95)      &  0.56 \, (0.93)   &  0.56 \, (0.92)  \\  \hline
   $jl \to h$ (virtual $W$) $\; \;$    &  $\; \; 6 \! \times \! 10^{-4}$  (0.012) $\; \;$ 
                                       &  $\; \; 9 \! \times \! 10^{-5}$  ($7.3 \! \times \! 10^{-3}$) $\; \;$   
                                       &  $\; \; 1.4 \! \times \! 10^{-4}$ ($4.8 \! \times \! 10^{-3}$) $\; \;$ \\
   $jl \to h$ (real $W$)         &  0.044 \, (0.96) &\   0.024 \, (0.95)    &\   0.017 \, (0.94) \\ 
   $t_{\rm semilep} \to h$                    &  0.043 \, (0.89) &\   0.04 \, (0.85)    &\   0.033 \, (0.83) 
\end{tabular} 
\caption{\it Tag rates for 200 GeV boosted Higgs bosons decaying to semileptonic $WW$ and mistag rates for QCD jets with an embedded lepton.  (The numbers in parentheses are tag rates for mini-isolation alone.)}
\label{tab:eff_200h_semi}
\end{table}

We choose an operating point with Higgs-jet efficiency very similar to the fully hadronic tag, namely about 50\% across the entire $p_T$ range.  The mistag rates for jets with embedded leptons via virtual $W$'s in hadron decays are extremely small, usually $O(10^{-3})$ or smaller.  Mini-isolation alone would yield percent or smaller mistag rates (consistent with~\cite{Rehermann:2010vq}), but the addition of multibody kinematic cuts is also significant.  In our subsequent $Z'$ searches in the next section, we find that backgrounds of this type are always subleading.  The remaining important sources of fake semileptonic diboson-jets, namely real $W$ emissions directly off of light quarks or in top quark decay, exhibit 1--10\% mistag rates.

\section{$Z' \to Zh$ Search}
\label{sec:results}

\begin{table}[tp]
 \centering
\begin{tabular}{ l|c c c }
                    &  $\;\; WW^{(*)}\to l\nu l\nu \;\;$ & $\;\; WW^{(*)}\to l\nu q\bar q' \;\;$  &  $\;\; WW^{(*)}\to q\bar q' q\bar q' \;\;$  \\ \hline
$Z\to l^+l^-$\      &  0.0032  &  0.020   &  0.030  \\
$Z\to \nu\bar\nu$\  &  0.0097  &  0.059   &  0.090  \\
$Z \to q\bar q$\    &  0.034   &  0.21    &  0.31   \\
\end{tabular} 
\caption{\it Branching fractions for $Z(WW^{(*)})$ into various final states (excepting taus).}
\label{tab:BRs}
\end{table}

With our substructure tools in place, we now demonstrate their utility in the context of the search for resonant $Zh$ production.  We focus on the case of a narrow TeV-scale $Z'$ produced in $q\bar q$ annihilation, and assume $m_h = 125$~GeV or 200~GeV.  We derive estimates of discovery sensitivity at the 14~TeV LHC, which can have appreciable $Z'$ production cross sections at truly multi-TeV masses, generating boosted Higgses with momenta in excess of a TeV.  We reserve an investigation of sensitivity/exclusion at the 7~or 8~TeV LHC for future work.\footnote{In particular, this may serve as an interesting context in which to find the Higgs in the first place (or other exotic scalars), especially if it is nonstandard.  However, this possibility must be carefully studied incorporating constraints from existing Higgs and new physics searches.}

Since we are ultimately dealing with a triboson signal, we face a large number of possible final states in which we can run a search.  To initially reduce the number of options, we neglect channels with tau-leptons.  But this still leaves us with nine possibilities, spanning a large range of branching fractions.  The situation is illustrated in Table~\ref{tab:BRs}.  As usual, channels with small branching fractions are typically cleaner because they have more leptonic activity.  However, even a channel with little background loses its utility if the total number of signal events is unobservably small.  Also, given the jet substructure tools outlined in the previous section, the benefits of trading off a quark for a lepton are not always as dramatic as we might first guess.

Our approach here will be to simply run a survey over most of the channels.  The one notable exception is the fully hadronic channel $Zh \to (q\bar q)(q\bar q' q\bar q')$, which faces a substantial background from dijet QCD and top pairs, even after exploiting the potential quintuply-resonant structure of the signal.  (Although for very high-mass searches, where the backgrounds have fallen off, this may nonetheless become competitive.)  This leaves us with eight exclusive channels, with varying degrees of leptonic, hadronic, and invisible activity.

For each of these eight channels, we simulate the signal and the most important backgrounds using leading-order matrix-elements in {\tt MadGraph} or {\tt PYTHIA}, supplemented with the leading-log virtuality-ordered parton shower in {\tt PYTHIA}.  We subsequently pass the events through the simple calorimeter model described in Appendix~\ref{sec:detector}.  We then cluster the event into quasi-hemispheric fat-jets of $R=1.5$, and decluster these as either $Z$-jets or diboson-jets.  The physics simulations and reconstructed object definitions are described in more detail in Appendix~\ref{sec:details}.

\begin{figure}[tp]
\begin{center}
\epsfxsize=0.44\textwidth\epsfbox{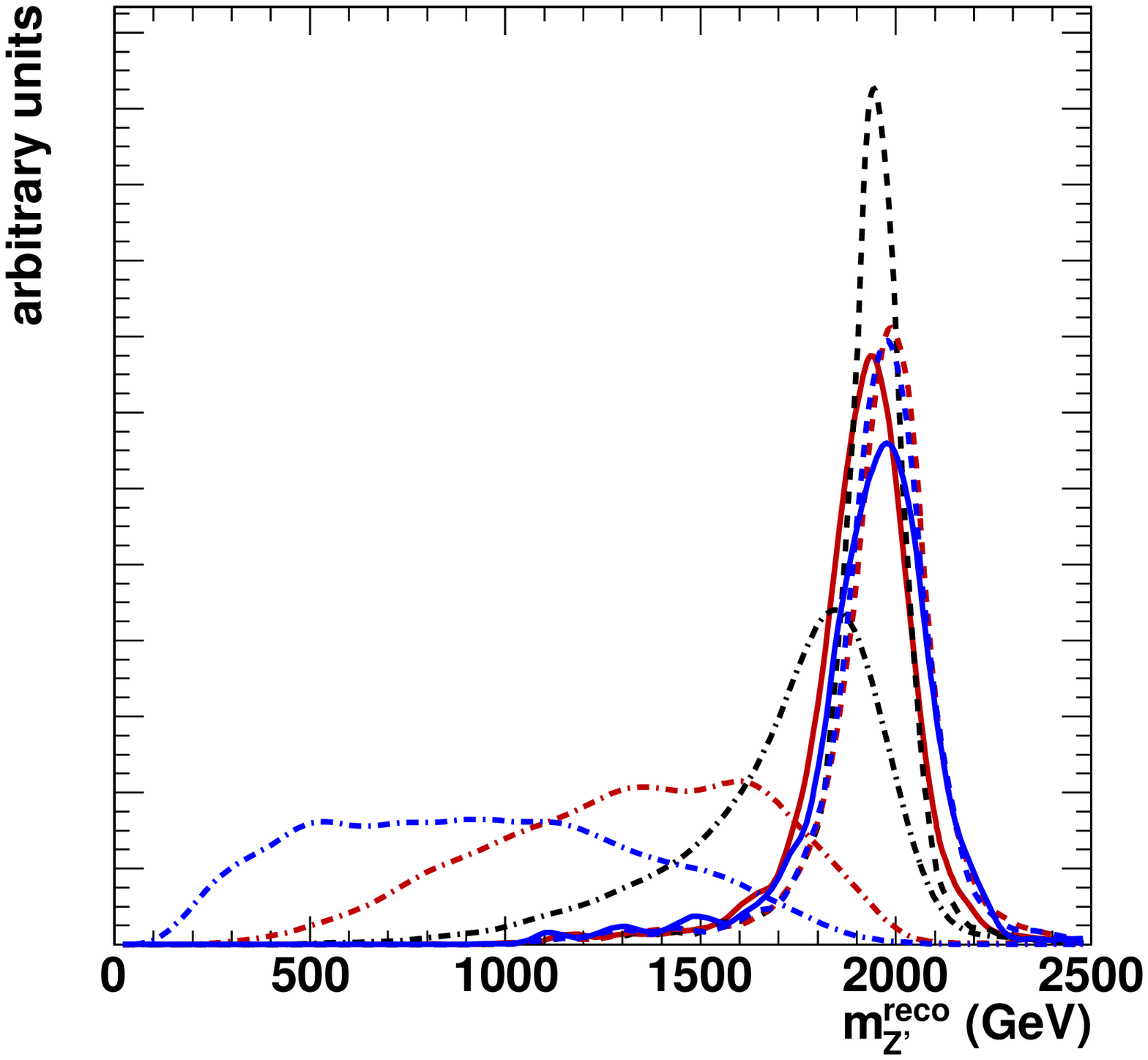}
\hspace{0.25in}
\epsfxsize=0.44\textwidth\epsfbox{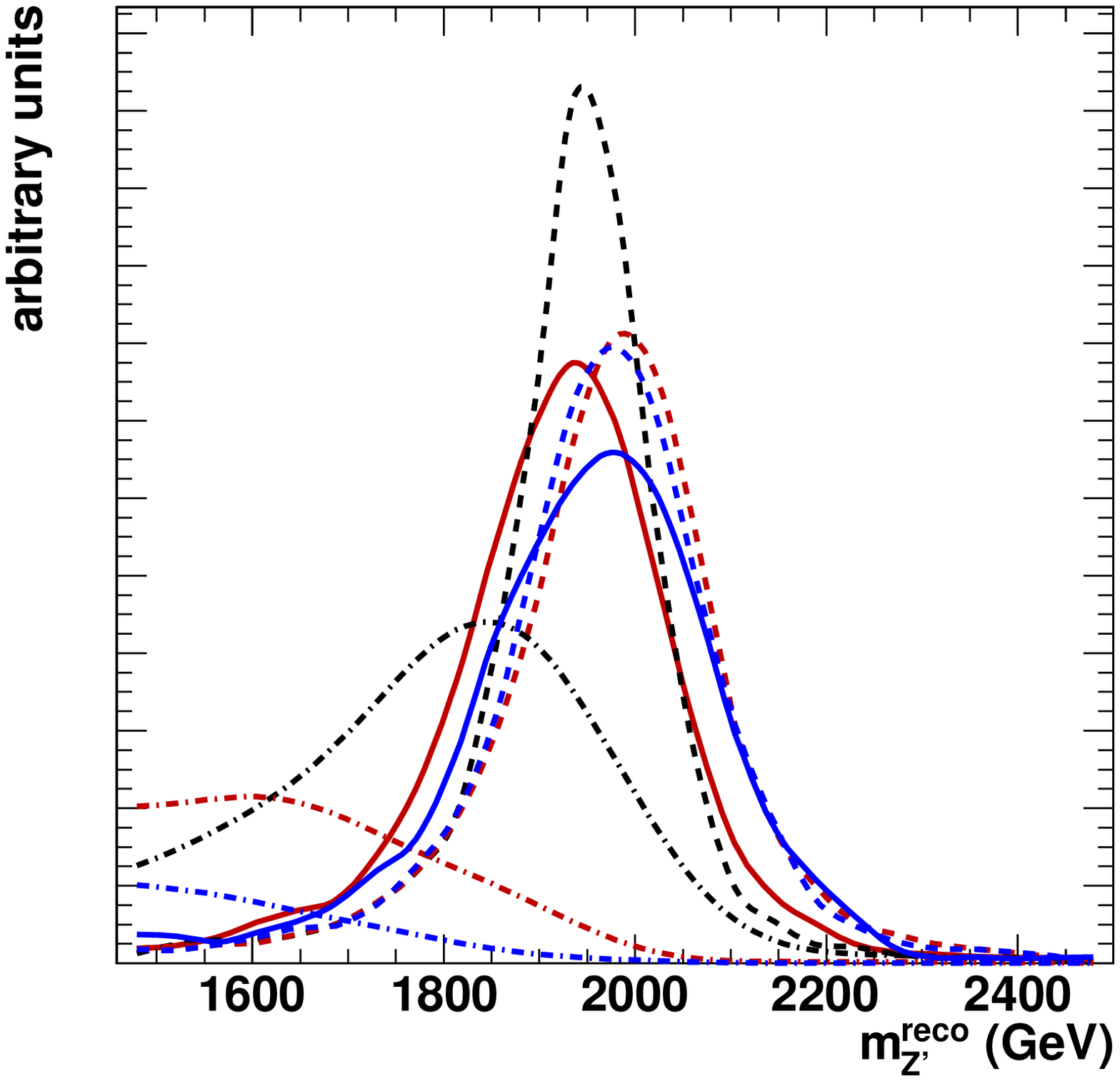}
\caption{\it Reconstructed $Z'$ lineshapes for a 125~GeV Higgs and $m_{Z'} = 2$~TeV in all eight of our analysis channels.  Colors indicate Higgs decay mode: $l\nu l\nu$ (blue), $l\nu q\bar q'$ (red), and $q\bar q' q\bar q'$ (black).  Dashing indicates $Z$ decay mode:  $l^+l^-$ (dashed), $\nu\bar\nu$ (dot-dashed), $q\bar q$ (solid).  The plot on the right is a zoom-in of the plot on the left.}
\label{fig:lineshape}
\end{center}
\end{figure}

The different channels face different issues related to energy/mass resolution and kinematic ambiguities, owing to different numbers of leptons versus jets and different numbers of neutrinos.  In cases where the $Z$ decays invisibly and/or the $WW^{(*)}$ system decays semi-invisibly, we default to a minimalistic reconstruction of a single effective neutrino:  the missing energy vector serves as the transverse momentum vector of the neutrino, and the neutrino's rapidity is set equal to that of the subsystem composed of the visible $WW^{(*)}$ decay products (subjets and/or leptons).  We then (semi-)reconstruct the event mass by adding all visible products and this effective neutrino.  In channels with a visible $Z$ decay, knowing the exact kinematics of the neutrinos is not so crucial for reconstructing the global mass of the event, and it is largely adequate to know the summed invisible $\vec{p}_T$ and approximate summed $p_z$.  In cases with an invisible $Z$ decay, we face an inevitable degradation of $Z'$ mass reconstruction, which gets worse as the $WW^{(*)}$ system becomes more invisible.\footnote{However, it is worth pointing out that a $Z'$ produced in $q\bar q$ annihilation is polarized along the beam, and is forbidden to decay back down to a scalar plus (longitudinal) gauge boson along the beamline.  This leads to a bias for central production, and more of the event's energy going transverse.  Consequently, the degrading effect on the peak can be less of an issue than it is for, say, the transverse mass distribution of singly-produced leptonic $W$ bosons.}  We illustrate the quality of our $Z'$ mass reconstruction in all eight of our search channels in Fig.~\ref{fig:lineshape}.

We structure all of our searches using a simple ``cut and count'' approach.  Details of the cuts vary from channel-to-channel, but a handful of common requirements can be identified:
\begin{itemize}
\item A tagged diboson-jet recoiling against either a tagged hadronic $Z$-jet, a leptonic $Z$, or substantial \met.
\item A sizeable ratio between the $p_T$ of each side of the event to the reconstructed $Z'$ mass.
\item A reconstructed $Z'$ mass localized in a window centered on the signal.
\end{itemize}
The second requirement only applies to cases where the $Z$ is visible.  It removes regions of phase space with high-$\sqrt{\hat s}$ but low-$p_T$, characteristic of many QCD-induced processes with $t$-channel singularities.  Usually, after the first two requirements, the signal appears as a localized excess on top of a smoothly-falling background distribution of reconstructed event mass.

After defining our search regions, we characterize the expected significance of the signal by taking the common $S/\sqrt{B}$ approximation, and determine the signal cross section required to achieve 5$\sigma$.  Since this becomes a very poor approximation for $O$(1) number of events, and since even nearly-vanishing physics backgrounds may in reality be supplemented by instrumental backgrounds, we introduce a statistical ``regulator'' by enforcing a floor of $B = 4$ events in each channel.  This means, for example, that even for channels with essentially zero expected background events according to our simple estimates, we require at least 10 signal events to claim 5$\sigma$ significance.

While sometimes one channel may clearly dominate the discovery reach, there is often much to be gained by forming a statistical combination.  This is especially true for the present situation, where the full signal appears distributed across a large number of exclusive analyses.  We perform our combination by forming a summed event count, with each channel weighted by $S/B$.  The final result is equivalent to adding together the significance of all channels in quadrature.

\begin{figure}[tp]
\begin{center}
\epsfxsize=0.44\textwidth\epsfbox{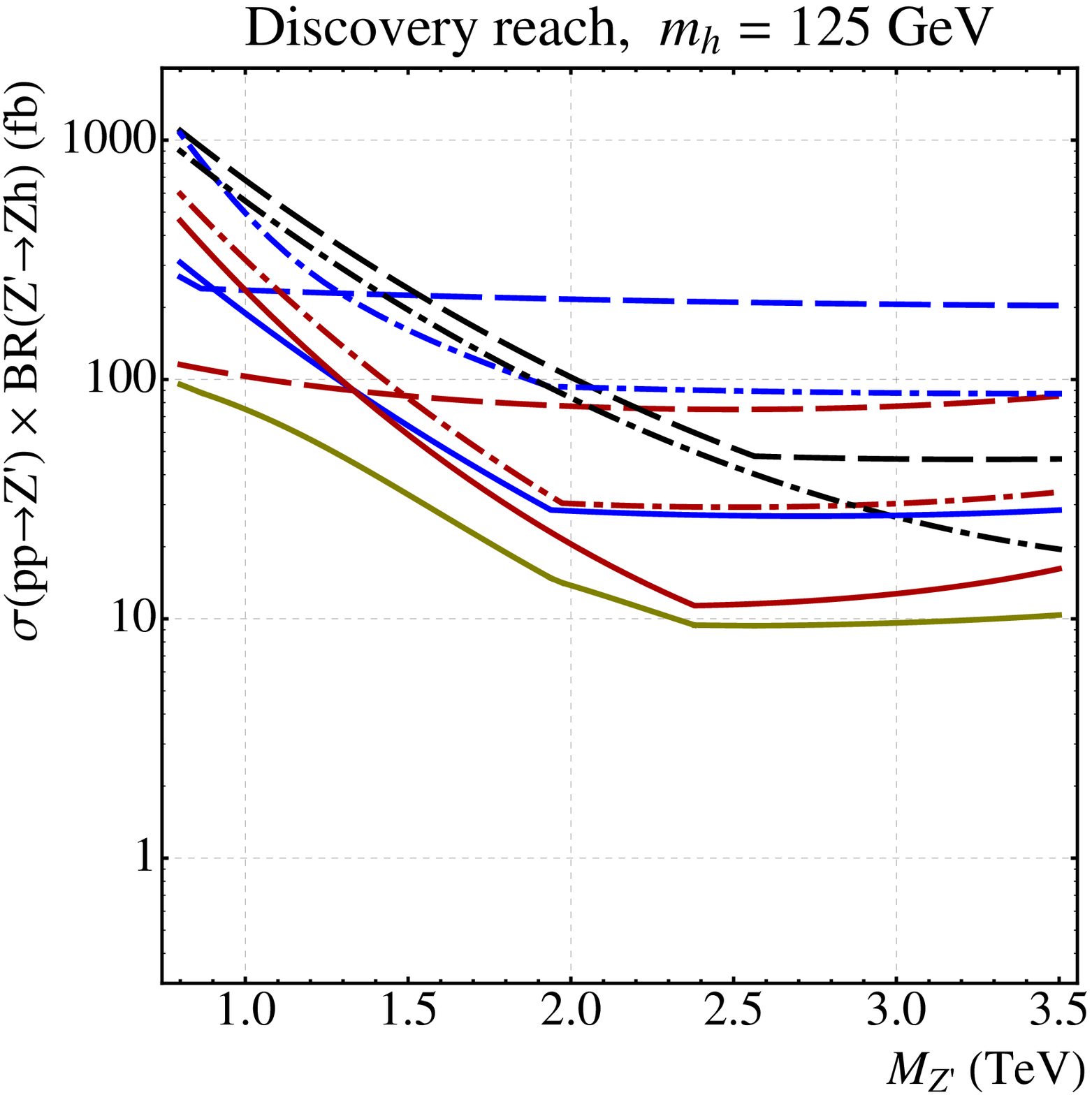}
\hspace{0.25in}
\epsfxsize=0.44\textwidth\epsfbox{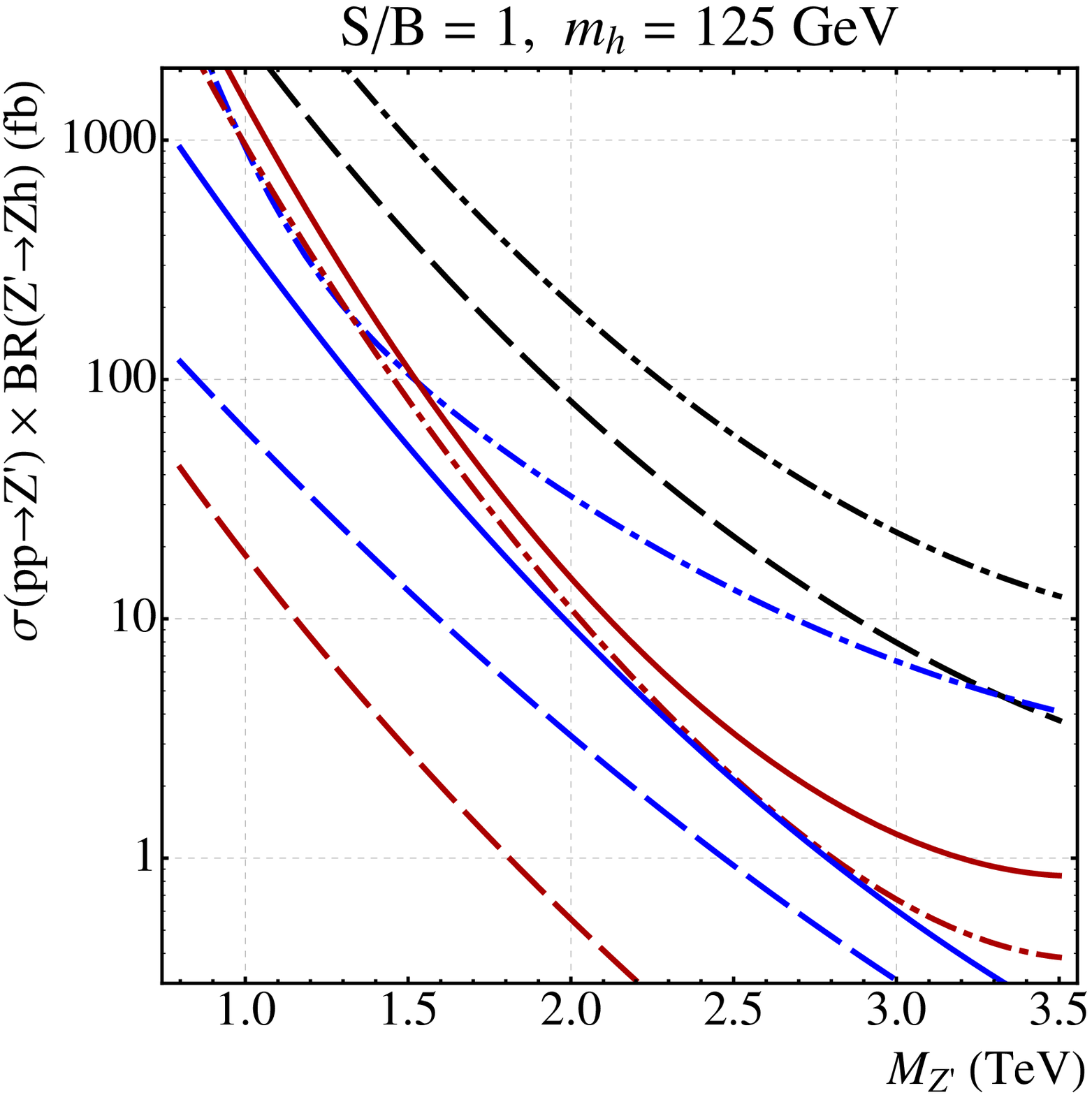}
\caption{\it Discovery reach with 100~fb$^{-1}$ (left) and $\sigma\times BR$ at which $S/B=1$ (right), for a 125~GeV Higgs.  Colors indicate Higgs decay mode: $l\nu l\nu$ (blue), $l\nu q\bar q'$ (red), and $q\bar q' q\bar q'$ (black).  Dashing indicates $Z$ decay mode:  $l^+l^-$ (dashed), $\nu\bar\nu$ (dot-dashed), $q\bar q$ (solid).  Dark yellow is a simple statistical combination.}
\label{fig:reach125}
\end{center}
\end{figure}

\begin{figure}[tp]
\begin{center}
\epsfxsize=0.44\textwidth\epsfbox{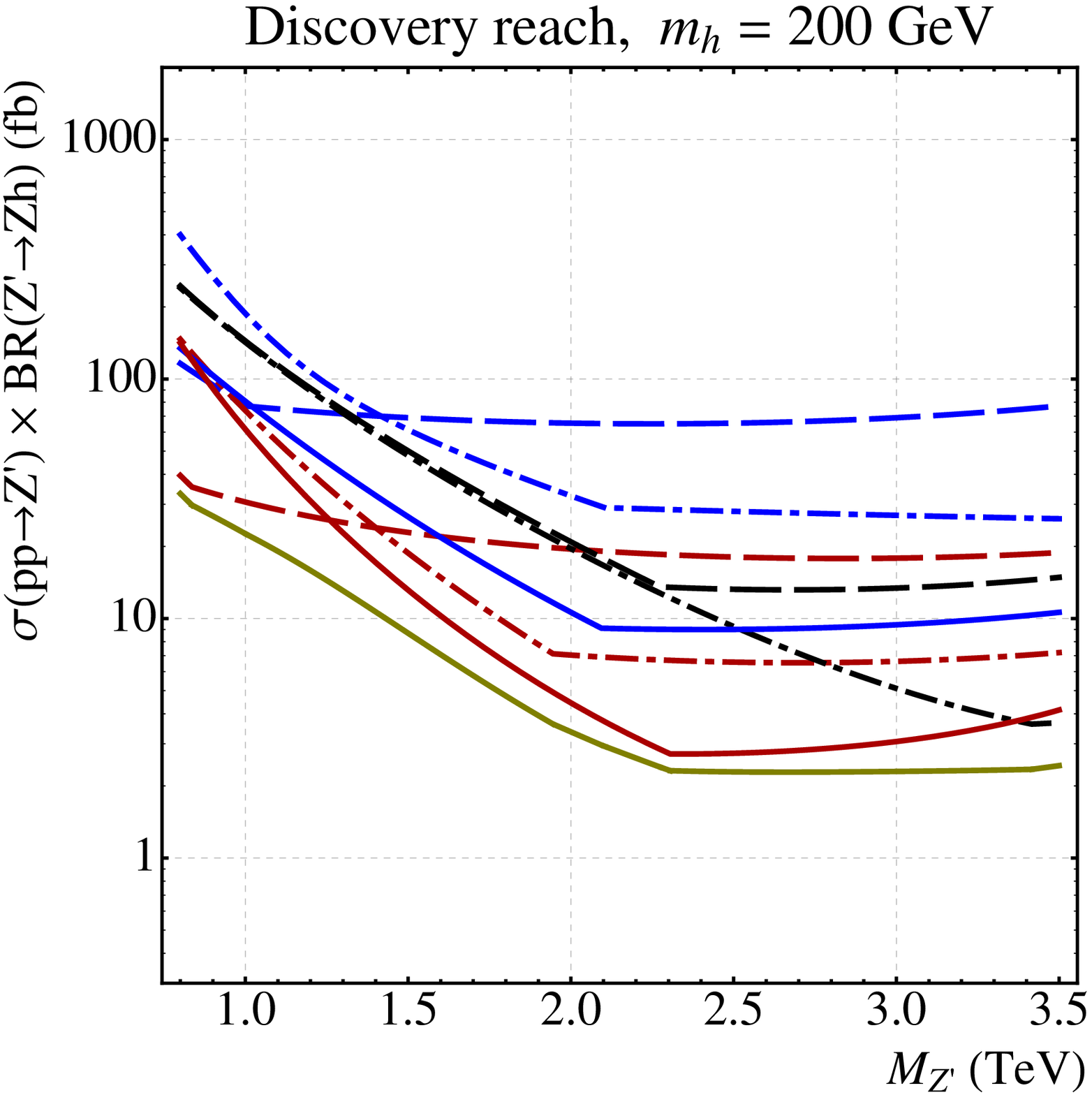}
\hspace{0.25in}
\epsfxsize=0.44\textwidth\epsfbox{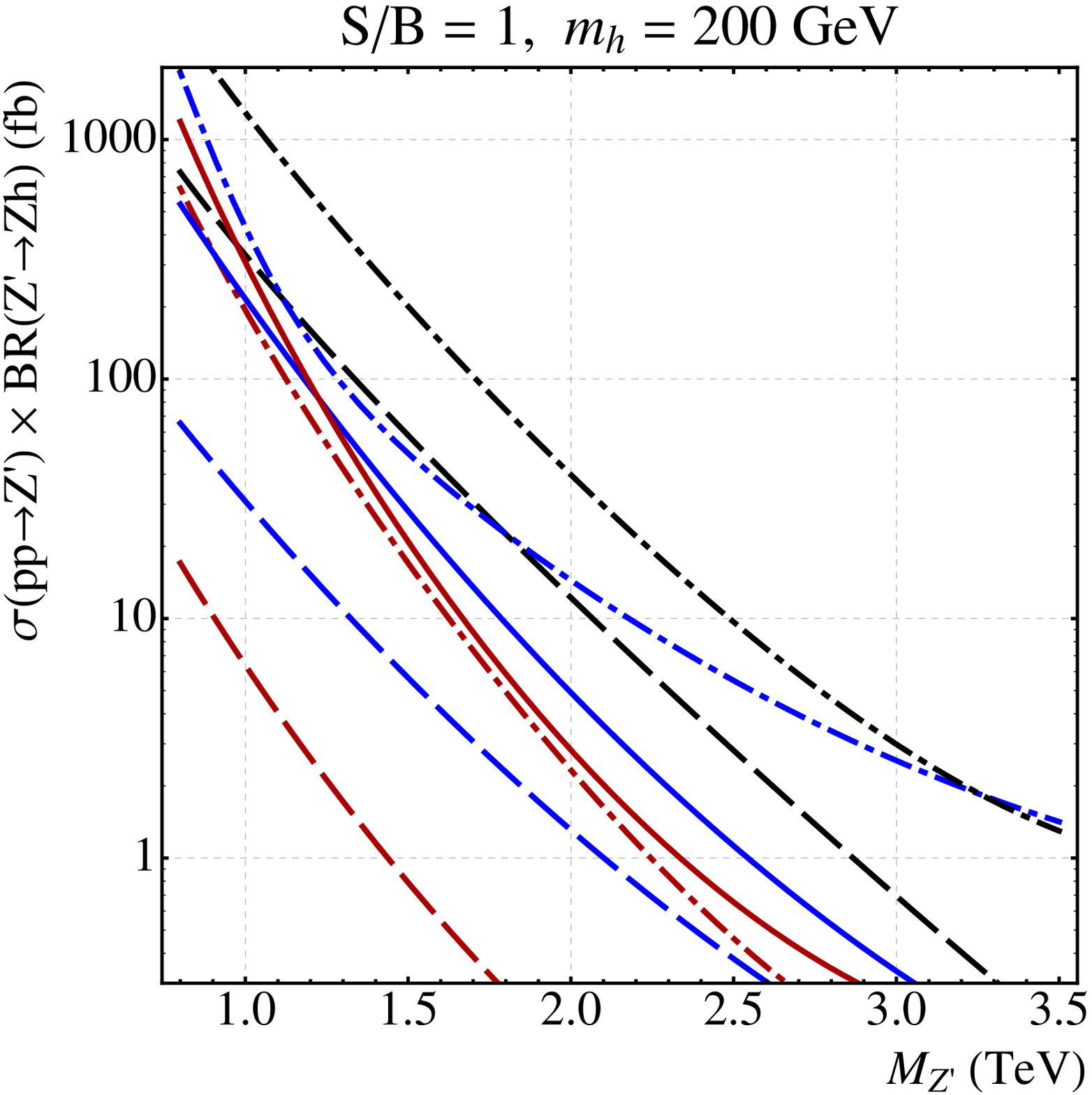}
\caption{\it Discovery reach with 100~fb$^{-1}$ (left) and $\sigma\times BR$ at which $S/B=1$ (right), for a 200~GeV Higgs.  Colors indicate Higgs decay mode: $l\nu l\nu$ (blue), $l\nu q\bar q'$ (red), and $q\bar q' q\bar q'$ (black).  Dashing indicates $Z$ decay mode:  $l^+l^-$ (dashed), $\nu\bar\nu$ (dot-dashed), $q\bar q$ (solid).  Dark yellow is a simple statistical combination.}
\label{fig:reach200}
\end{center}
\end{figure}

We present our final search reach estimates in Figs.~\ref{fig:reach125} and~\ref{fig:reach200} for the 125~GeV and 200~GeV Higgs, respectively.  (The $WW^{(*)}$ branching fractions are 22\% and 74\%.)  The plots show the minimum $\sigma(pp \to Z')\times BR(Z'\to Zh)$ required for a 5$\sigma$ discovery after a 100~fb$^{-1}$ run of the 14~TeV LHC.  They are supplemented by plots of the $\sigma\times BR$ required for $S/B=1$, to give some sense of the relative size of the backgrounds and the degree to which we might worry about systematic errors.\footnote{The smallest $S/B$'s which we encounter at a claimed discovery limit are $O(1/10)$, corresponding to an excess of a few hundred signal events on top of a few thousand background events.}  (Here we do not explicitly attempt to estimate systematic errors associated with background shape/normalization.)

The most powerful individual search channel depends on the $Z'$ mass and LHC luminosity, as well as on the precise cuts which have been applied.  However, we can infer some general features regarding how the channels compare with one another given on our own set of choices.  Channels with large numbers of leptons, such as the four-lepton $Zh \to (l^+l^-)(l\nu l\nu)$, have very little background but are so rate-limited that they cannot compete at high masses.  Indeed, it is the mostly-hadronic modes $(q\bar q)(l\nu q\bar q')$ and $(\nu\bar\nu)(q\bar q' q\bar q')$ that dominate the search reach above 3~TeV for 100~fb$^{-1}$.  This is greatly facilitated by the jet substructure methods which we have applied (hadronic $Z$-tag, semileptonic and hadronic diboson-tags), which heavily suppress the otherwise sizeable backgrounds.\footnote{The channel $Zh \to (q\bar q)(l\nu q\bar q')$ was considered before in~\cite{Agashe:2007ki}, which found similar model reach at high mass without application of jet substructure techniques.  We note that those results capitalized on very strict kinematic cuts without application of showering or energy resolution effects, and did not include top backgrounds.  Still, the question of how to best trade off between substructure cuts and more traditional kinematic cuts remains a nontrivial one.}  At masses below 3~TeV, the situation is more mixed, but we find that leaders are $(q\bar q)(l\nu q\bar q')$, $(q\bar q)(l\nu l\nu)$, and $(\nu\bar\nu)(l\nu q\bar q)$.  Below about $1.2$~TeV, $(l^+l^-)(l\nu q\bar q')$ becomes dominant.  Each of these channels benefits in some way from two-body hadronic substructure, and most benefit from the full semileptonic diboson-jet tagger.

Of course, given the typically $O(1)$ differences in discovery reach that we have found, a more complete experimental analysis would be required to make any solid claims of relative performance.  Still, it is interesting that jet substructure makes so many of these channels competitive, whereas a more conservative approach might discard many of the hadronic options in favor of modes with more leptons.

\begin{figure}[tp]
\begin{center}
\epsfxsize=0.44\textwidth\epsfbox{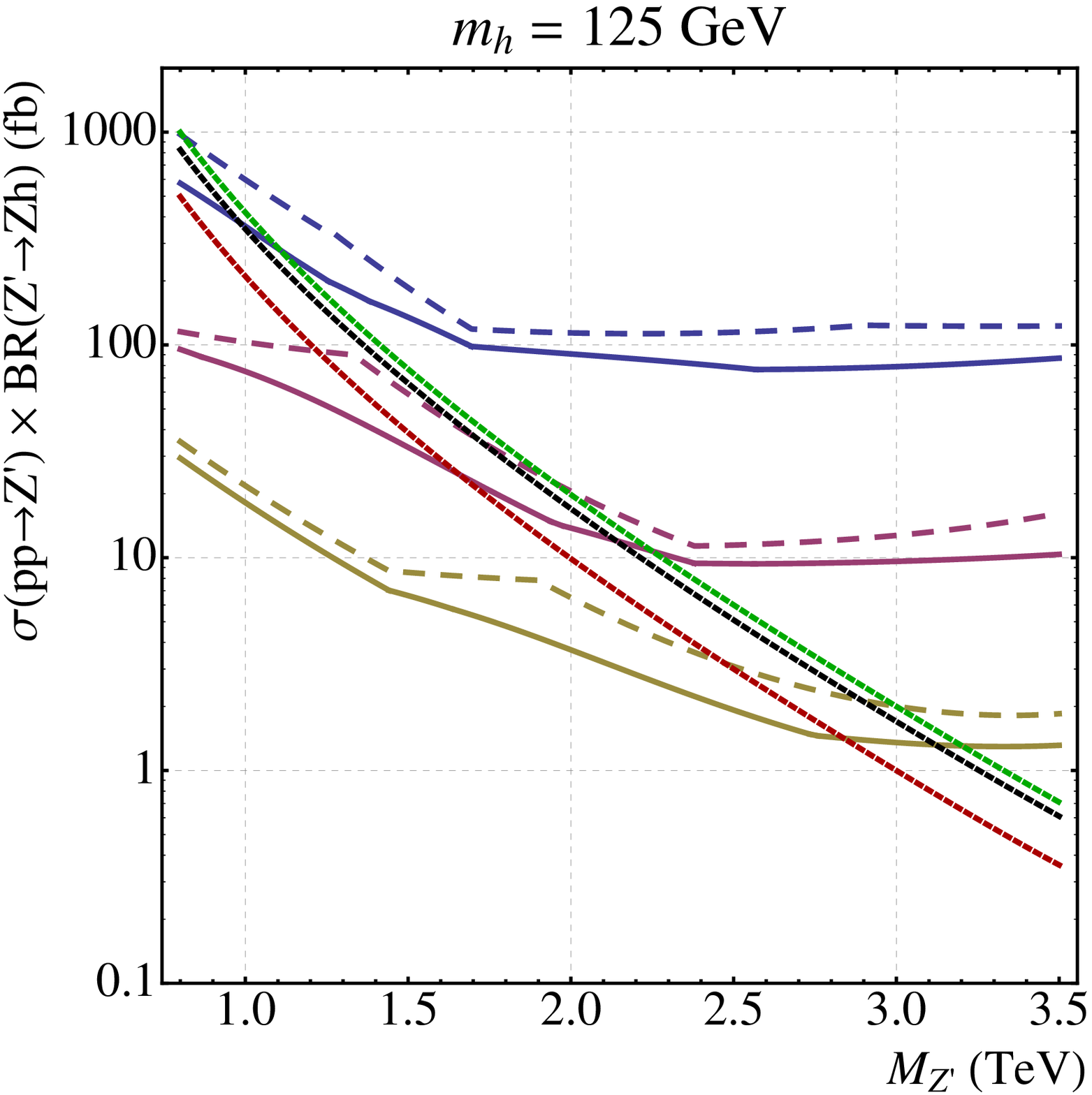}
\hspace{0.25in}
\epsfxsize=0.44\textwidth\epsfbox{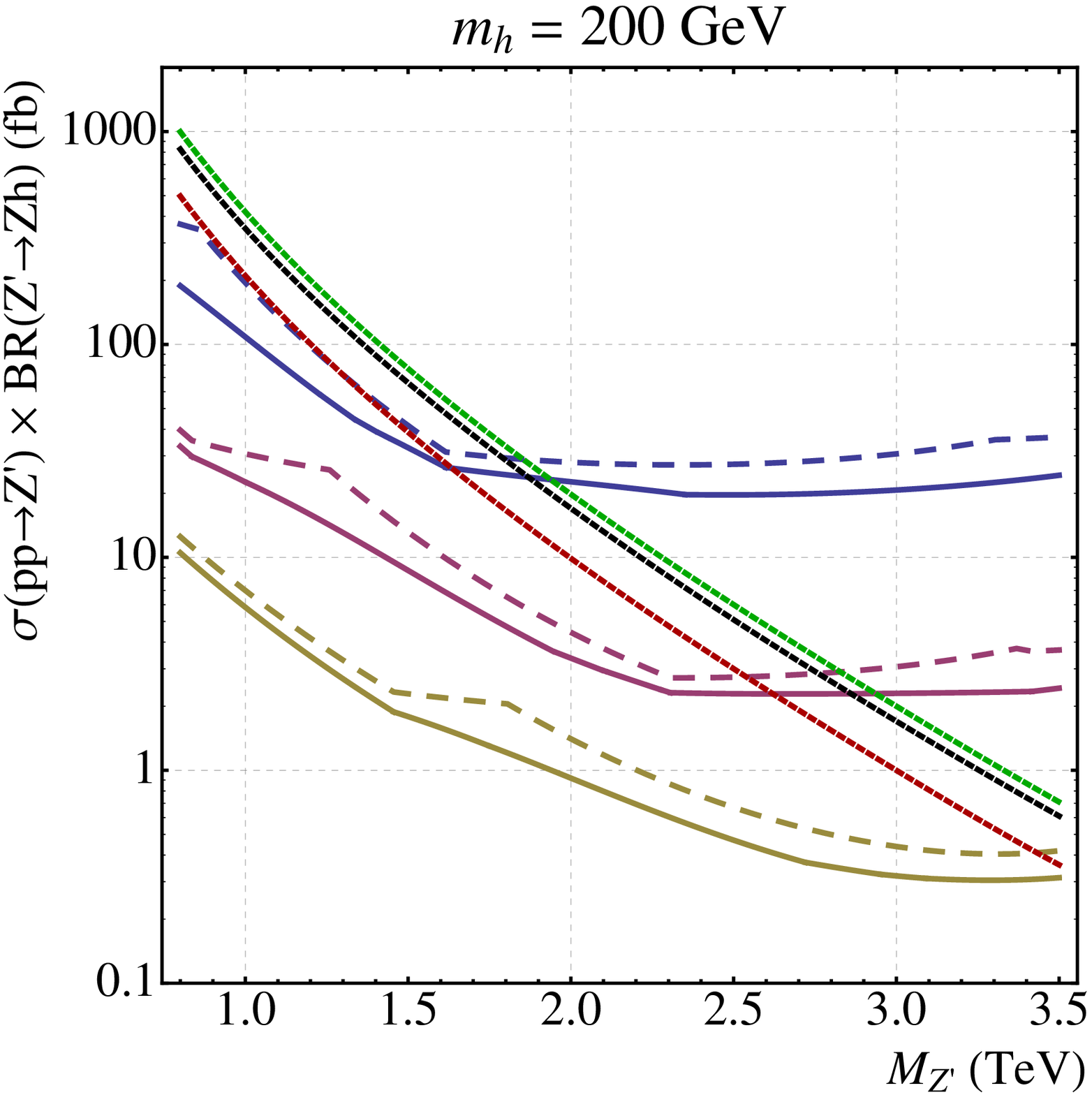}
\caption{\it Discovery reach for different luminosities for a 125~GeV Higgs (left) and a 200~GeV Higgs (right), superimposed with $Z'$ model cross section curves.  The displayed luminosities are 10~fb$^{-1}$ (blue), 100~fb$^{-1}$ (purple), and 1000~fb$^{-1}$ (dark yellow).  Solid lines are a simple statistical combination of all modes, and dashed lines are the best individual mode for a given $Z'$ mass.  Models are indicated by short-dashed lines, and include a $Z$-sequential boson (green), the warped KK $Z'$ bosons of~\cite{Agashe:2003zs} (black), and either the $\eta$ boson of $E_6$ or a $T_R^3$ gauge boson (red).}
\label{fig:reachVsL}
\end{center}
\end{figure}

In Fig.~\ref{fig:reachVsL}, we show how the discovery reach evolves with luminosity, picking either the best channel for a given mass point or performing the combination of channels.  We also show four baseline models with sizeable $\sigma\times BR$ into $Zh$:  a sequential $Z'$, the warped KK $Z'$ bosons of~\cite{Agashe:2003zs}, the $\eta$ boson of the $E_6$ unified model, and the $T_R^3$ boson of a left-right symmetric model with $SU(2)_R \to U(1)$ broken at a high scale (reviewed in~\cite{Langacker:2008yv}).\footnote{We define the $Z'$-Higgs couplings in the sequential $Z'$ model by replacing $Z_\mu \to Z_\mu + (m_Z/m_{Z'})^2 Z'_\mu$ in all $Z$-Higgs couplings.  This yields $BR(Z'\to Zh) \simeq 3$\%, which is comparable to the $BR$ into electrons (or muons).  The warped KK $Z'$ is a set of three neutral bosons with variable $BR$s to $Zh$, described in detail in~\cite{Agashe:2007ki}.  For the $E_6$ and $T_R^3$ models, we assume that a single ``up-type'' Higgs dominates electroweak symmetry-breaking, in which case the $Zh$ (electron/muon) $BR$s are respectively 5.3\% (3.3\%) and 2.3\% (4.5\%).   Dilepton resonance searches at the LHC constrain the sequential, $E_6$, and $T_R^3$ bosons to be above about 2~TeV~\cite{ATLASdilep,CMSdilep}.  Electroweak precision constraints suggest that the KK, $E_6$, and $T_R^3$ bosons are heavier than about 3~TeV.}  (The last two have rates that are within 10\% of one another, and they are represented by a single line.)  Many of these models have roughly democratic couplings to Standard Model fermions and to the Higgs doublet, whereas the KK bosons have small couplings to the former interplaying with large couplings to the latter.  For the 200~GeV Higgs, which is excluded if it has Standard Model-like couplings~\cite{ATLASHiggsCombined,CMSHiggsCombined}, we can instead take as a baseline a fermiophobic Higgs (as in, e.g.,~\cite{Diaz:1994pk}) with the same $U(1)'$ charge as its SM-like counterpart.  While there are many ways to hide the Higgs boson, this assumption keeps $BR(Z'\to Zh)$ and $BR(h\to WW)$ largely unchanged, and allows the simplest comparison to the 125~GeV analysis.  Still, any realistic fermiophobic model would also require an additional source of electroweak symmetry breaking to give mass to the top quark and other fermions, and consequently there will be some degree of suppression of $Z'\to Zh$ since the fermiophobic Higgs cannot saturate EWSB.

Within the context of this diverse handful of $Z'$ models, we see that the 14 TeV LHC has good discovery prospects for masses well above a TeV.  Even for the 125~GeV case, where the $WW^*$ decay mode is subleading, models with masses in the 1.5--2.5~TeV range are visible with $O$(100~fb$^{-1}$) luminosity.  For the 200~GeV case, the mass reach potentially goes up to about 2.5--3.0~TeV, modulo the necessity of additional model-dependence to hide the Higgs from direct searches.  We have also indicated what might be achieved with a 1000~fb$^{-1}$ luminosity upgrade:  3~TeV or higher mass reach for the 125~GeV Higgs, and better than 3.5~TeV mass reach for the 200~GeV Higgs.

If the Higgs indeed resides at 125~GeV, and is largely standard in its couplings, then the $h\to WW^*$ decay rate is smaller than $b\bar b$ by about a factor of 3.  The study of~\cite{Katz:2010mr} suggests that a search for $Z' \to Zh \to Z(b\bar b)$ would be more sensitive, typically by an $O(1)$ factor in $\sigma\times BR$.  (The comparison is nontrivial, as the present diboson study can utilize the $Z\to q\bar q$ decays, whereas if the Higgs also decays to jets, even $b$-jets, the corresponding channels are probably highly background-contaminated.)  Nonetheless, even if these estimates are correct, a corroborating observation of $Z' \to Zh \to Z(WW^*)$ would be extremely useful.  It would not only serve to verify the existence of the $Z'$, but would improve confidence that the $b\bar b$ resonance into which it decays is really the Higgs.  Our present results suggest that this is not only possible, but that both discoveries may occur on a similar timescale.  Of course, there is still a chance that Nature will surprise us, either by yielding up a Higgs which is actually dominated by $WW^{(*)}$, or by providing additional scalar states with diboson decay modes.

\section{Conclusions}
\label{sec:conclusion}


In this paper we have considered how to discriminate boosted $WW^{(*)}/ZZ^{(*)}$ ``diboson-jet'' systems utilizing jet substructure techniques, focusing on the cases of $h \to WW^{*}$ for a 125~GeV Higgs and $h \to WW$ for a 200~GeV Higgs.  For fully hadronic decays, we demonstrated the feasibility of a dedicated tagger by identifying the hardest splittings in the diboson-jet, filtering out diffuse radiation, and exploiting the kinematic distributions of the resulting four hard subjets.  For tag rates of approximately 50\%, we achieve mistag rates between $0.5$--5\%.  For semileptonic decays, we showed that mini-isolation and two-body hadronic substructure could be combined to powerfully reject QCD jets with embedded leptons from heavy flavor decays, as well as jets with $W$-strahlung emission and semileptonic top-jets.  Again working at 50\% tag rate, we reject the former at the $10^{-4}$ level, and find $O$(5\%) mistags for jets with real $W$s.  For both $WW^{(*)}$ decay modes, in order to keep the tag/mistag rates stable as we exceed $p_T \simeq $ TeV, and as the angular scales shrink to the size of individual hadronic calorimeter elements, we have investigated the possibility of a novel ECAL/HCAL hybridization.

As a simple application of these diboson-jet taggers, we have outlined searches for a TeV scale $Z' \to Zh$ resonance at the 14~TeV LHC using the various $Z(WW^{(*)})$ final-state channels.  We found that channels with one or more hadronic $W$ and/or $Z$ decays are more sensitive than the exclusively leptonic channels.  In particular, the leaders are $Zh \to (q\bar q)(l\nu q\bar q')$, $(\nu\bar\nu)(l\nu q\bar q')$, $(\nu\bar\nu)(q\bar q' q\bar q')$, $(q\bar q)(l\nu l\nu)$, and, at masses below about a TeV, $(l^+l^-)(l\nu q\bar q')$.  Only one of these channels uses the rare but highly distinctive fully leptonic $WW^{(*)} \to (l\nu l\nu)$ decay, and, as we scan up in $Z'$ masses, ultimately runs out of statistics faster than channels that capitalize on fully hadronic or semileptonic diboson-jet tagging.  Taken together, our results suggest that resonances in the truly multi-TeV mass range should be discoverable via the diboson decay modes of the Higgs, and that searches for these resonances would be substantially facilitated and expanded by jet substructure techniques.

While this demonstrates the long-term LHC reach using extremely collimated diboson systems, we emphasize that our techniques become useful even at momenta only a few times higher than the Higgs mass itself, which may be accessible to the current 7~and 8~TeV LHC.  Understanding $Z'$ sensitivity at these energies, or even the prospects for first discovery of a nonstandard Higgs in $Z'$ decay, is an interesting open topic.

Although we have exclusively studied the specific decay sequence $Z' \to Zh \to Z(WW^{(*)})$ for simplicity, we note that many of the same techniques will also apply to $W' \to Wh \to W(WW^{(*)})$, as well as to the subdominant Higgs decay mode $h \to ZZ^{(*)}$.  The latter case in particular was considered in the six-lepton final state in \cite{Barger:2009xg}, albeit with very limited reach in $Z'$ mass due to the very small total branching fraction.  Adding in modes with partial or fully hadronic decays of the boosted $ZZ^{(*)}$ system is a straightforward generalization of our treatment of $WW^{(*)}$, and indeed the latter would be picked up for free anyway.  (This would increase statistics for our 125~GeV and 200~GeV analyses by about 10\% and 30\%, respectively, though we did not explicitly include them.)  Incorporating the $ZZ^{(*)} \to l^+l^-q\bar q$ modes could offer some interesting supplementary search channels, such as $Zh \to Z(ZZ^{(*)}) \to (q\bar q)(l^+l^-q\bar q)$, which has similar topology to $Zh \to Z(WW^{(*)}) \to (q\bar q)(l\nu q\bar q')$.  Still, the small total rates make these modes less attractive for a multi-TeV resonance search, where statistics is a major limiting factor.

Diboson-jet tagging can have a wider application beyond searches involving boosted Higgses.  Indeed, if the Higgs is near 125~GeV, as suggested by the LHC data, then our $Z'$ search could also be accomplished using the dominant Higgs decay mode $h \to b\bar b$~\cite{Katz:2010mr} (or the subdominant but highly distinctive ditau mode~\cite{Katz:2010iq}).  However, many models of new physics also predict scalars beyond the Higgs, which can dominantly decay into $WW^{(*)}$ through mixing \cite{Fox:2011qc}.  A particularly relevant example in the context of renormalizable $Z'$ models would be the Higgs(es) of the $U(1)'$ sector.  Other types of boosted multi-jet cascades from new physics can also benefit from our treatment, even if the intermediate state is not $WW^{(*)}/ZZ^{(*)}$.

There are also many other ways to produce boosted Higgses, such as in heavy top partner decays or due to strong EWSB effects in weak boson scattering~\cite{Contino:2010mh}.  Having diboson-jet tagging available as a tool for identifying these processes could greatly improve sensitivity, especially in final states with multiple boosted Higgses where we can include both $b\bar b$ and $WW^*$ decay modes.

\acknowledgments{We are grateful to Andrey Katz for collaboration in the early phase of this work.  MS was supported in part by the DoE under grant No.\ DE-FG-02-92ER40704.  CS and BT were supported by DoE grant No.\ DE-FG-02-91ER40676. BT was also supported by NSF-PHY-0969510 (LHC Theory Initiative).}

\appendix

\section{Analysis Details}
\label{sec:details}

We generate $Z' \to Zh \to Z(WW^{(*)})$ signal samples, keeping all spin correlations, using \MadGraph\ {\tt v4.5.1}~\cite{Alwall:2007st} interfaced with \PYTHIA\ (virtuality-ordered shower).  For background samples, we use {\tt MadGraph5} {\tt v1.3.30}~\cite{Alwall:2011uj}.  All processes are evaluated at leading order, and without pileup.\footnote{In our previous paper~\cite{Katz:2010mr}, we found that the effects of pileup were easily controllable.  For the present paper, we have also studied the application of trimming~\cite{Krohn:2009th}.  We assume that most charged pileup particles can be eliminated by vertexing, leaving over only photons and neutral hadrons.  If we subsequently trim each event by pre-clustering into $R = 0.2$ anti-$k_T$ jets~\cite{Cacciari:2008gp} and throwing away jets with $p_T < $~a few~GeV, then we find that the influence of pileup on our reconstructions is largely eliminated.}  We do not attempt to use matrix elements to model the multibody distribution of partons inside a jet, except in cases involving a heavy particle decay (such as a top quark).  So, for example, for a QCD jet faking a hadronic diboson-jet, the jet is generated by a single hard parton, and its substructure is generated entirely by parton showering.  As noted in Section~\ref{sec:jetsub}, there is sensitivity to the exact showering model when dealing with hadronic substructure.  In such cases we have erred on the conservative side by rescaling the background rate to match what would be expected from the {\tt HERWIG} shower.

We pass the showered and hadronized samples through the detector simulation described in Appendix~\ref{sec:detector}, and proceed to reconstruct jets, leptons and \met.  Finally, we apply substructure methods and analysis cuts.

For leptons, we smear the energies by 2\% for electrons and by (5\%)$\times\sqrt{E/{\rm TeV}}$ for muons.  Subsequently, we only consider leptons with $p_T > 30$ GeV and  $|\eta| < 2.5$.  We make a first pass over the event, searching for electrons and muons which are isolated in a traditional sense:  scalar-summing the $p_T$ of all photons and hadrons within an $\eta$--$\phi$ cone of $R=0.4$ around the lepton, $p_T(l)/(p_T(l)+p_T({\rm cone})) > 0.9$.  (Note that we do not include other nearby leptons in the cone.)  Amongst these, any opposite-sign same-flavor pairs with a mass $m(l^+l^-) = [81,101]$ GeV are added together and treated as leptonic $Z$ bosons.  If, after $Z$ boson clustering, there are no other isolated leptons in the event, then we check whether there are any non-isolated leptons, and keep the hardest one if it is mini-isolated as defined in Section~\ref{sec:semilep}.  All remaining leptons are considered ``hadrons'' for jet clustering, below.

We then cluster all calorimeter cells (and non-isolated leptons) into jets using the Cambridge/Aachen algorithm with $R = 1.5$, as implemented in \FastJet\ {\tt v2.4.2}~\cite{Cacciari:2005hq}.  These quasi-hemispheric fat-jets serve as the input into our jet substructure algorithms.  We only consider events where the highest-$p_T$ fat-jet has $|\eta| < 1.5$.  In events with no reconstructed leptons (but possibly a leptonic $Z$), we decluster the hardest jet using the four-body hadronic diboson-tagger and require that it pass all of our internal phase space cuts on the four-subjet mass and subjet pairwise masses described in Section~\ref{sec:had}.  All other jets we decluster into two subjets with the BDRS mass-drop method~\cite{Butterworth:2008iy}.  We fractionally smear all subjet energies by $5\;({\rm GeV}/p_T) \, \oplus \,  0.5\sqrt{{\rm GeV}/p_T} \, \oplus \, 0.04$.

We reconstruct \met~by balancing the $\vec{p}_T$ of all leptons and the subjets from the leading two decomposed fat-jets.

The final analysis channel of an event depends on a small set of baseline criteria applied to the subjets, leptons, and \met.  In all cases with one lepton, we attempt to match it with the hardest subjet-pair within $\Delta R < 1.5$ to form a semi-leptonic diboson candidate.  We then classify the event based on the activity on the opposite side of the detector ($\Delta\phi > \pi/2$ away from the vector-summed subjets and lepton):  a leptonic $Z$, a $Z$-candidate subjet-pair with $m = [76,106]$~GeV, or substantial \met.  (A jet is only considered as a hadronic $Z$-candidate if it is also the hardest jet in the event.)  If the recoiling $Z$ is visible, we apply the semileptonic diboson-tag incorporating the lepton's sister neutrino, whereas if the $Z$ is invisible we only utilize the lepton and matched subjets.  The full semileptonic tag criteria are described in Section~\ref{sec:semilep}.  In cases with two leptons, which serve as dileptonic diboson candidates, we require that the leptons be within $\Delta R < 1.5$ of one another.  Again, we determine the final channel by checking whether the opposite side of the detector is consistent with leptonic $Z$, hadronic $Z$, or invisible $Z$.  Finally, if there is no lepton, then the hardest jet should be a hadronic diboson candidate, and we only consider cases where it is associated with a leptonic $Z$ or \met\ comparable to the jet $p_T$.  This defines our eight analysis channels.

All channels under consideration, except for hadronic $WW^{(*)}$ with leptonic $Z$, contain neutrinos.  In such cases, in order to facilitate approximate Higgs and $Z'$ reconstruction, we define a single effective neutrino by using the \met\ vector as the neutrino's $p_T$ vector.  We then set the neutrino's rapidity equal to that of the visible diboson decay products.  In general, the reconstructed $Z'$ mass, $m_{Z'}^{\rm reco}$, is defined as the invariant mass of the four-vector sum of the visible diboson decay products, visible $Z$ decay products (if any), and the effective neutrino.


\begin{table}[p]
\centering
\begin{tabular}{|c|c|} \hline
  \multicolumn{2}{|c|}{Fully hadronic $WW^{(*)}$ associated with} \\ \hline
  {\rm Leptonic}\ $Z$  & {\rm Invisible}\ $Z$  \\ \hline \hline
  $p_T(h) > m_{Z'}^{\rm reco}/3$ &   \\ \hline
  \quad \quad $m_{Z'} - 15\% < m^{\rm reco}_{Z'} < m_{Z'} + 15\%$ \quad \quad & \quad \quad $m_{Z'} - 30\% < m^{\rm reco}_{Z'} < m_{Z'} + 10\%$ \quad \quad \\  \hline
\end{tabular}
\caption{\it Kinematic cuts imposed for the analysis of $Z' \to Zh \to Z(q\bar q' q\bar q')$.}
\label{tab:hadronicWW}
\end{table}

\begin{table}[tp]
 \centering
\begin{tabular}{|c|c|c|} \hline
  \multicolumn{3}{|c|}{Semileptonic $WW^{(*)}$ associated with} \\ \hline
  {\rm Leptonic}\ $Z$  &  {\rm Hadronic}\ $Z$  &   {\rm Invisible}\ $Z$ \\ \hline \hline
  \multicolumn{2}{|c|}{ $p_T(j_W j_W) > 0.1 \times p_T(Z)$ } &   $p_T(j_W j_W) > 0.3 \times \mathmet$ \\
  \multicolumn{2}{|c|}{ $p_T(Z) > m_{Z'}^{\rm reco}/3$ } &   $p_T(j_2) < 0.3 \times p_T(j_W j_W)$ \\ \hline
  $m_{Z'} - 15\% < m^{\rm reco}_{Z'} < m_{Z'} + 15\%$\  & $m_{Z'} - 10\% < m^{\rm reco}_{Z'} <  m_{Z'} + 10\%$  &\ $m^{\rm reco}_{Z'} >   m_{Z'}/2$ \\ \hline
\end{tabular}
\caption{\it Kinematic cuts imposed for the analysis of $Z' \to Zh \to Z(l\nu q\bar q')$.  $j_2$ represents the sum of subjets from the second-hardest fat-jet.}
\label{tab:semilepWW}
\end{table}

\begin{table}[tp]
\centering
\begin{tabular}{|c|c|c|} \hline
  \multicolumn{3}{|c|}{Dileptonic $WW^{(*)}$ associated with} \\ \hline
  \;\; {\rm Leptonic}\ $Z$ \;\; & {\rm Hadronic}\ $Z$ & {\rm Invisible}\ $Z$  \\ \hline \hline
  \multicolumn{2}{|c|}{ $h$(125): \hspace{0.3cm} $m(l_1^+ l_2^-\nu) < 135$} & \; \; $10 < m(l_1^+ l_2^-) < 75$ \; \; \\
  \multicolumn{2}{|c|}{ $h$(200): \hspace{0.3cm} $m(l_1^+ l_2^-\nu) < 210$} & \; \; $10 < m(l_1^+ l_2^-) < 150$ \; \; \\ \hline 
  \multicolumn{2}{|c|}{ $p_T(Z) > m_{Z'}^{\rm reco}/3$}  &  $p_T(j_1) < 0.3 \, \times $ \met\   \\ \hline
  \multicolumn{2}{|c|}{ $m_{Z'} - 10\% < m^{\rm reco}_{Z'} < m_{Z'} + 10\%$}  &  $m^{\rm reco}_{Z'} > 0.3 \times m_{Z'}$ \\ \hline
\end{tabular}
\caption{\it Kinematic cuts imposed for the analysis of $Z' \to Zh \to Z(l\nu l\nu)$.  (All units are in GeV.)  $j_1$ represents the sum of subjets from the hardest fat-jet.}
\label{tab:dilepWW}
\end{table}


We subsequently apply cuts to define a signal region, and determine the efficiency for the signal and the cross section for each background.\footnote{When defining signal efficiencies, we do not allow cross-talk between the different decay channels and their corresponding search channels.  For example, decays involving taus can also produce (mini-)isolated leptons, and can sometimes get picked up by one of our leptonic search channels.  However, when determining the efficiency for a given leptonic channel, we do not include signal events with taus.}  The cuts are given in Tables~\ref{tab:hadronicWW} through~\ref{tab:dilepWW}.  We note that none of these cuts has been rigorously optimized in the sense of a computer scan, but they have all simply been tuned by eye to reject as much background as possible while keeping $O(1)$ signal efficiency.


\begin{table}[p]
\centering
{\scriptsize
\begin{tabular}{|l|c|c|c|c|c|c|}  \hline
 & \multicolumn{2}{|c|}{$m_{Z'} = 1 \ {\rm TeV}$} & \multicolumn{2}{|c|}{$m_{Z'} = 2 \ {\rm TeV}$} & \multicolumn{2}{|c|}{$m_{Z'} = 3 \ {\rm TeV}$} \\ \cline{2-7}
                   & 125 GeV    & 200 GeV  & 125 GeV & 200 GeV  & 125 GeV & 200 GeV\\ \hline \hline
Signal Eff.                   & 21\%   & 17\%  &  28\%  & 30\%  & 31\%  & 32\%  \\ \hline
$qg \to qZ$      & 2.0   & 0.54  &\ 6.9 $\times 10^{-2}$ \ &\ 3.3 $\times 10^{-2}$ \  &\ 7.0 $\times 10^{-3}$ \ &\ 1.8 $\times 10^{-3}$ \ \\
$q\bar q \to gZ$ & 0.66  & 0.33  &\ 3.6 $\times 10^{-2}$ \ &\ 2.4 $\times 10^{-2}$ \  &\ 4.4 $\times 10^{-3}$ \ &\ 1.7 $\times 10^{-3}$ \ \\ \hline \hline
 subleading & \multicolumn{6}{|l|}{$Ztb$, $Zt\bar t$, diboson, triboson, continuum $Zh$} \\ \hline
\end{tabular}
}
\caption{\it Signal efficiency and background cross sections (in fb) after all cuts in the $Zh \to (l^+l^-)(q\bar q' q\bar q')$ channel.  Signal efficiency does not incorporate the channel's total $BR$ of 0.0066 (0.022) for 125~GeV (200~GeV) Higgs.}
\label{tab:lljjjjEff}
\end{table}

\begin{table}[p]
 \centering
{\scriptsize
\begin{tabular}{|l|c|c|c|c|c|c|}  \hline
 & \multicolumn{2}{|c|}{$m_{Z'} = 1 \ {\rm TeV}$} & \multicolumn{2}{|c|}{$m_{Z'} = 2 \ {\rm TeV}$} & \multicolumn{2}{|c|}{$m_{Z'} = 3 \ {\rm TeV}$} \\ \cline{2-7}
                 & 125 GeV & 200 GeV     & 125 GeV                   & 200 GeV  & 125 GeV & 200 GeV\\ \hline \hline
Signal Eff.                    & 28\%    & 24\%  &  37\%  &  39\% &  41\% &  43\% \\ \hline
$qg \to q(Z/W)$        &  27   & 16   & 1.0   & 0.79  & 0.13                  & 4.7 $\times 10^{-2}$ \\
$q\bar q \to g(Z/W)$   &  11   & 5.0  & 0.50  & 0.25  & 5.1 $\times 10^{-2}$   & 3.8 $\times 10^{-2}$ \\ \hline \hline
 subleading & \multicolumn{6}{|l|}{$Ztb$, $Zt\bar t$, single-top, diboson, triboson, continuum $Zh$} \\ \hline
\end{tabular}
}
\caption{\it Signal efficiency and background cross sections (in fb) after all cuts in the $Zh \to (\nu\bar\nu)(q\bar q' q\bar q')$ channel.  Signal efficiency does not incorporate the channel's total $BR$ of 0.020 (0.067) for 125~GeV (200~GeV) Higgs.}
\label{tab:vvjjjjEff}
\end{table}

\begin{table}[p]
 \centering
{\scriptsize
\begin{tabular}{|l|c|c|c|c|c|c|}  \hline
 & \multicolumn{2}{|c|}{$m_{Z'} = 1 \ {\rm TeV}$} & \multicolumn{2}{|c|}{$m_{Z'} = 2 \ {\rm TeV}$} & \multicolumn{2}{|c|}{$m_{Z'} = 3 \ {\rm TeV}$} \\ \cline{2-7}
                 & 125 GeV & 200 GeV     & 125 GeV                   & 200 GeV  & 125 GeV & 200 GeV\\ \hline \hline
Signal Eff.               & 21\%  & 21\%  &  28\%  &  33\% & 28\% &  36\%\\ \hline
$ZWj$                  & 1.3 $\times 10^{-2}$  & 1.3 $\times 10^{-2}$    & 5.7 $\times 10^{-4}$  & 4.2 $\times 10^{-4}$   & 3.9 $\times 10^{-5}$   & 3.9 $\times 10^{-5}$  \\
$Zt\bar t$             & 4.2 $\times 10^{-3}$  & 6.1 $\times 10^{-3}$   & 9.7 $\times 10^{-5}$  & 1.6 $\times 10^{-4}$   & 7.0 $\times 10^{-6}$   & 1.2 $\times 10^{-5}$  \\
triboson               & 6.8 $\times 10^{-4}$  & 1.7 $\times 10^{-3}$   & 5.2 $\times 10^{-5}$  & 1.8 $\times 10^{-4}$   & 5.7 $\times 10^{-6}$   & 2.4 $\times 10^{-5}$  \\ \hline \hline
 subleading & \multicolumn{6}{|l|}{$Zj$, $Ztb$, continuum $Zh$} \\ \hline
\end{tabular}
}
\caption{\it Signal efficiency and background cross sections (in fb) after all cuts in the $Zh \to (l^+l^-)(l\nu q\bar q')$ channel.  Signal efficiency does not incorporate the channel's total $BR$ of 0.0044 (0.015) for 125~GeV (200~GeV) Higgs.}
\label{tab:lllvjjEff}
\end{table}

\begin{table}[p]
 \centering
{\scriptsize
\begin{tabular}{|l|c|c|c|c|c|c|}  \hline
 & \multicolumn{2}{|c|}{$m_{Z'} = 1 \ {\rm TeV}$} & \multicolumn{2}{|c|}{$m_{Z'} = 2 \ {\rm TeV}$} & \multicolumn{2}{|c|}{$m_{Z'} = 3 \ {\rm TeV}$} \\ \cline{2-7}
                 & 125 GeV & 200 GeV     & 125 GeV                   & 200 GeV  & 125 GeV & 200 GeV\\ \hline \hline
Signal Eff.      & 18\%    & 20\%  &  25\%  &  32\% &  25\%  &  34\% \\ \hline
$l\nu j$   & 1.3   &  1.1  & 2.1 $\times 10^{-2}$  & 2.5 $\times 10^{-2}$  & 8.5 $\times 10^{-4}$ & 9.5 $\times 10^{-4}$ \\
$t\bar t$  & 0.81  & 0.48  & 1.0 $\times 10^{-2}$  & 4.5 $\times 10^{-3}$  & 1.1 $\times 10^{-3}$ & 6.2 $\times 10^{-4}$ \\
$ZWj$      & 0.15  & 0.14  & 5.4 $\times 10^{-3}$  & 3.6 $\times 10^{-3}$  & 2.7 $\times 10^{-4}$ & 7.3 $\times 10^{-4}$ \\ \hline \hline
 subleading & \multicolumn{6}{|l|}{$Zj$, diboson, triboson, single-top, $Ztb$, $Zt\bar t$, continuum $Zh$}  \\ \hline
\end{tabular}
}
\caption{\it Signal efficiency and background cross sections (in fb) after all cuts in the $Zh \to (\nu\bar\nu)(l\nu q\bar q')$ channel.  Signal efficiency does not incorporate the channel's total $BR$ of 0.013 (0.044) for 125~GeV (200~GeV) Higgs.}
\label{tab:vvlvjjEff}
\end{table}

\begin{table}[tp]
 \centering
{\scriptsize
\begin{tabular}{|l|c|c|c|c|c|c|}  \hline
 & \multicolumn{2}{|c|}{$m_{Z'} = 1 \ {\rm TeV}$} & \multicolumn{2}{|c|}{$m_{Z'} = 2 \ {\rm TeV}$} & \multicolumn{2}{|c|}{$m_{Z'} = 3 \ {\rm TeV}$} \\ \cline{2-7}
                        & 125 GeV  & 200 GeV   & 125 GeV  & 200 GeV  & 125 GeV & 200 GeV\\ \hline \hline
Signal Eff.             & 14\%  & 13\%  &  19\% &  23\% &  17\%  & 21\% \\ \hline
$t\bar t$       & 5.1   & 4.2   & 7.2 $\times 10^{-2}$  & 4.7 $\times 10^{-2}$  & 4.9 $\times 10^{-3}$  & 3.4 $\times 10^{-3}$ \\
$Wjj$           & 4.2   & 2.0   & 5.8 $\times 10^{-2}$  & 5.4 $\times 10^{-2}$  & 1.1 $\times 10^{-2}$  & 4.8 $\times 10^{-3}$ \\ \hline \hline
 subleading & \multicolumn{6}{|l|}{QCD dijets, $(W/Z)j$, diboson+jet, triboson, single-top, continuum $Zh$}  \\ \hline
\end{tabular}
}
\caption{\it Signal efficiency and background cross sections (in fb) after all cuts in the $Zh \to (q\bar q)(l\nu q\bar q')$ channel.  Signal efficiency does not incorporate the channel's total $BR$ of 0.046 (0.16) for 125~GeV (200~GeV) Higgs.}
\label{tab:jjlvjjEff}
\end{table}

\begin{table}[tp]
 \centering
{\scriptsize
\begin{tabular}{|l|c|c|c|c|c|c|}  \hline
& \multicolumn{2}{|c|}{$m_{Z'} = 1 \ {\rm TeV}$} & \multicolumn{2}{|c|}{$m_{Z'} = 2 \ {\rm TeV}$} & \multicolumn{2}{|c|}{$m_{Z'} = 3 \ {\rm TeV}$} \\ 
\cline{2-7}
                 & 125 GeV & 200 GeV     & 125 GeV                   & 200 GeV  & 125 GeV & 200 GeV\\ \hline \hline

Signal Eff.            
& 35\%
& 32\%
& 46\% 
& 42\%
& 48\% 
& 41\% \\ \hline

$l^+l^-j$         
& 0.44 
& 0.71 
& 1.4 $\times 10^{-2}$ 
& 1.9 $\times 10^{-2}$ 
& 1.0 $\times 10^{-3}$ 
& 1.4 $\times 10^{-3}$ \\

$\tau^+ \tau^- j$ 
& 0.33 
& 0.35
& 1.1 $\times 10^{-2}$ 
& 1.1 $\times 10^{-2}$ 
& 7.4 $\times 10^{-4}$ 
& 7.8 $\times 10^{-4}$ \\ 

$t \bar{t} j$ 
& 0.24
& 0.66
& 6.6 $\times 10^{-3}$ 
& 2.0 $\times 10^{-2}$ 
& 3.9 $\times 10^{-4}$ 
& 1.1 $\times 10^{-3}$ \\

$Zl^+l^- \to jjl^+l^-$
& 5.9 $\times 10^{-2}$
& 8.4 $\times 10^{-2}$
& 2.4 $\times 10^{-3}$
& 3.0 $\times 10^{-3}$
& 1.8 $\times 10^{-4}$
& 2.7 $\times 10^{-4}$ \\

$W^+W^-j$         
& 2.8 $\times 10^{-2}$ 
& 7.1 $\times 10^{-2}$ 
& 1.5 $\times 10^{-3}$ 
& 3.4 $\times 10^{-3}$ 
& 1.1 $\times 10^{-4}$ 
& 3.0 $\times 10^{-4}$ \\ \hline \hline

subleading                            
& \multicolumn{6}{l|}{triboson, continuum $Zh$}  \\ \hline

\end{tabular}
}
\caption{\it Signal efficiency and background cross sections (in fb) after all cuts in the $Zh \to (q\bar q)(l\nu l\nu)$ channel.  Signal efficiency does not incorporate the channel's total $BR$ of 0.0075 (0.025) for 125~GeV (200~GeV) Higgs.}
\label{tab:jjlvlvEff}
\end{table}

\begin{table}[tp]
 \centering
{\scriptsize
\begin{tabular}{|l|c|c|c|c|c|c|}  \hline
& \multicolumn{2}{|c|}{$m_{Z'} = 1 \ {\rm TeV}$} & \multicolumn{2}{|c|}{$m_{Z'} = 2 \ {\rm TeV}$} & \multicolumn{2}{|c|}{$m_{Z'} = 3 \ {\rm TeV}$} \\ 
\cline{2-7}
                 & 125 GeV & 200 GeV     & 125 GeV                   & 200 GeV  & 125 GeV & 200 GeV\\ \hline \hline

Signal Eff.            
& 55\%
& 50\%
& 60\% 
& 59\%
& 63\% 
& 56\% \\ \hline

$Zl^+l^- \to 4l$  incl. $\tau$'s
& 2.6 $\times 10^{-2}$
& 4.0 $\times 10^{-2}$
& 1.5 $\times 10^{-3}$
& 2.0 $\times 10^{-3}$
& 1.5 $\times 10^{-4}$
& 2.0
 $\times 10^{-4}$\\ \hline \hline

subleading                            
& \multicolumn{6}{l|}{$Z t \bar{t}$, triboson, continuum $Zh$}  \\ \hline

\end{tabular}
}
\caption{\it Signal efficiency and background cross sections (in fb) after all cuts in the $Zh \to (l^+l^-)(l\nu l\nu)$ channel.  Signal efficiency does not incorporate the channel's total $BR$ of 0.00070 (0.0024) for 125~GeV (200~GeV) Higgs.}
\label{tab:lllvlvEff}
\end{table}

Our final results appear in Tables~\ref{tab:lljjjjEff} through~\ref{tab:vvlvlvEff}.\footnote{To model the contribution of the $W$+jet background in the $(\nu\bar\nu)(q\bar q' q\bar q')$ channel, we have simply rescaled the $Z$+jet contribution by $1.5$.  This brings us into rough agreement with existing monojet searches (e.g.,~\cite{CMSmonojetRS}) without requiring us to explicitly model how well a leptonic $W$ can be vetoed.}  We determine the signal efficiencies and background cross sections for a trial set of $Z'$ masses:  1, 2, and 3~TeV.   (Many of the backgrounds that we checked are subleading, in the sense that together they add up to only $O(10\%)$ or less of the total.  For brevity, we do not explicitly list their cross sections.)   We extend these results to arbitrary $Z'$ masses by quadratic interpolation of the signal efficiencies and of the logarithms of the individual background cross sections.

\begin{table}[tp]
 \centering
{\scriptsize
\begin{tabular}{|l|c|c|c|c|c|c|}  \hline
& \multicolumn{2}{|c|}{$m_{Z'} = 1 \ {\rm TeV}$} & \multicolumn{2}{|c|}{$m_{Z'} = 2 \ {\rm TeV}$} & \multicolumn{2}{|c|}{$m_{Z'} = 3 \ {\rm TeV}$} \\ 
\cline{2-7}
                 & 125 GeV & 200 GeV     & 125 GeV                   & 200 GeV  & 125 GeV & 200 GeV\\ \hline \hline
Signal Eff.                      & 45\% & 43\% & 51\% & 48\% & 54\% & 52\% \\ \hline
$(\nu\bar\nu)(l^+l^-)$ incl. $\tau$'s & 0.22  & 0.22  & 2.6 $\times 10^{-2}$ & 2.9 $\times 10^{-2}$ & 6.1 $\times 10^{-3}$ & 6.3 $\times 10^{-3}$ \\
$(l\nu)(l^+l^-)$ incl. $\tau$'s       & 0.19  & 0.38  & 1.0 $\times 10^{-2}$ & 2.3 $\times 10^{-2}$ & 1.7 $\times 10^{-3}$ & 3.4 $\times 10^{-3}$ \\ 
$t\bar t$                             & 0.48  & 0.73  & $< 8 \times 10^{-4}$ & $< 8 \times 10^{-4}$ & $< 1 \times 10^{-4}$ & $< 1 \times 10^{-4}$ \\ \hline\hline
subleading                            & \multicolumn{6}{l|}{$(l\nu)(l\nu)$, triboson, continuum $Zh$}  \\ \hline
\end{tabular}
}
\caption{\it Signal efficiency and background cross sections (in fb) after all cuts in the $Zh \to (\nu\bar\nu)(l\nu l\nu)$ channel.  Signal efficiency does not incorporate the channel's total $BR$ of 0.0021 (0.0072) for 125~GeV (200~GeV) Higgs.}
\label{tab:vvlvlvEff}
\end{table}


\FloatBarrier

\section{Detector Model}
\label{sec:detector}


\begin{figure}[tp]
\begin{center}
\epsfxsize=0.44\textwidth\epsfbox{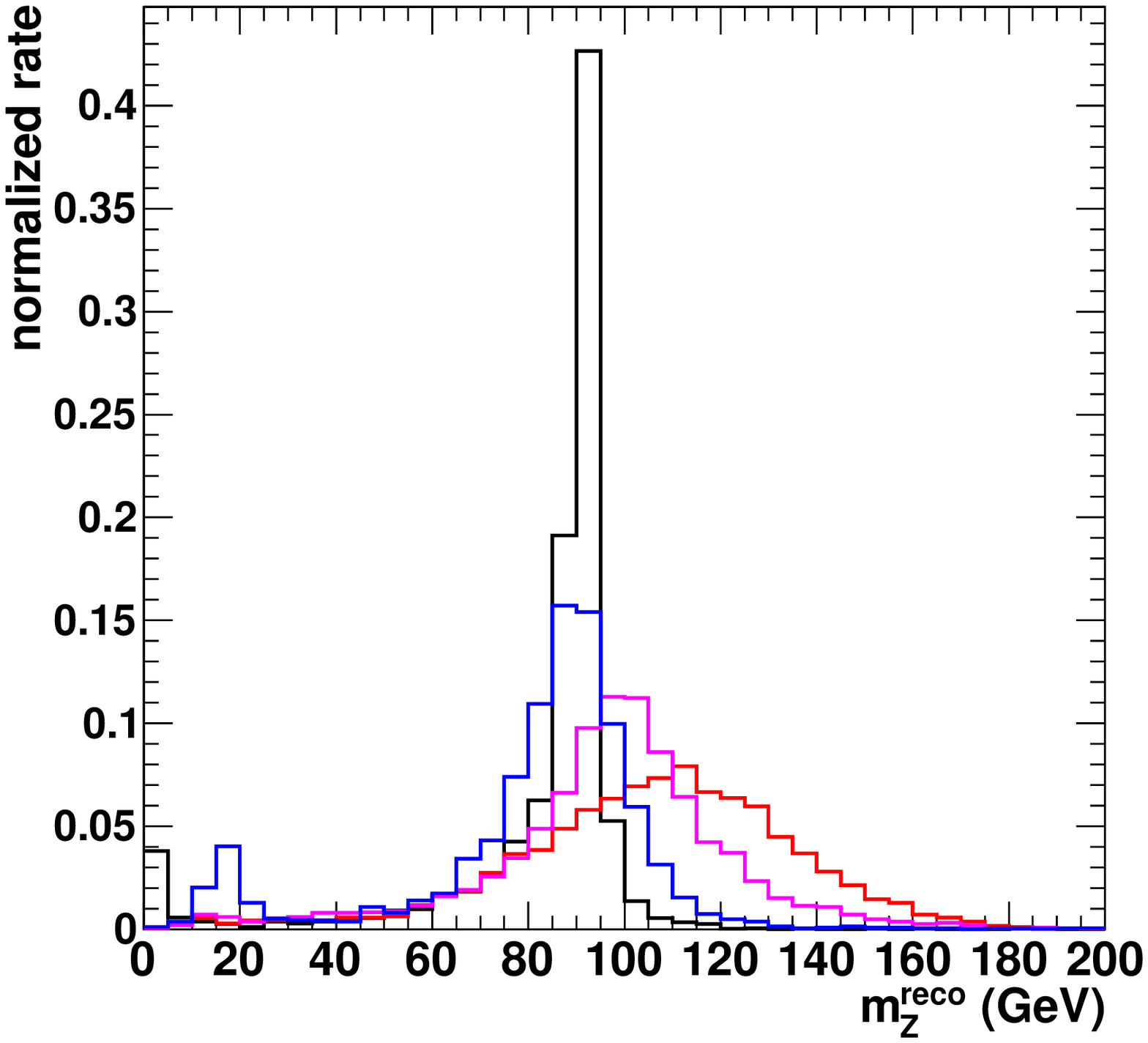}
\epsfxsize=0.44\textwidth\epsfbox{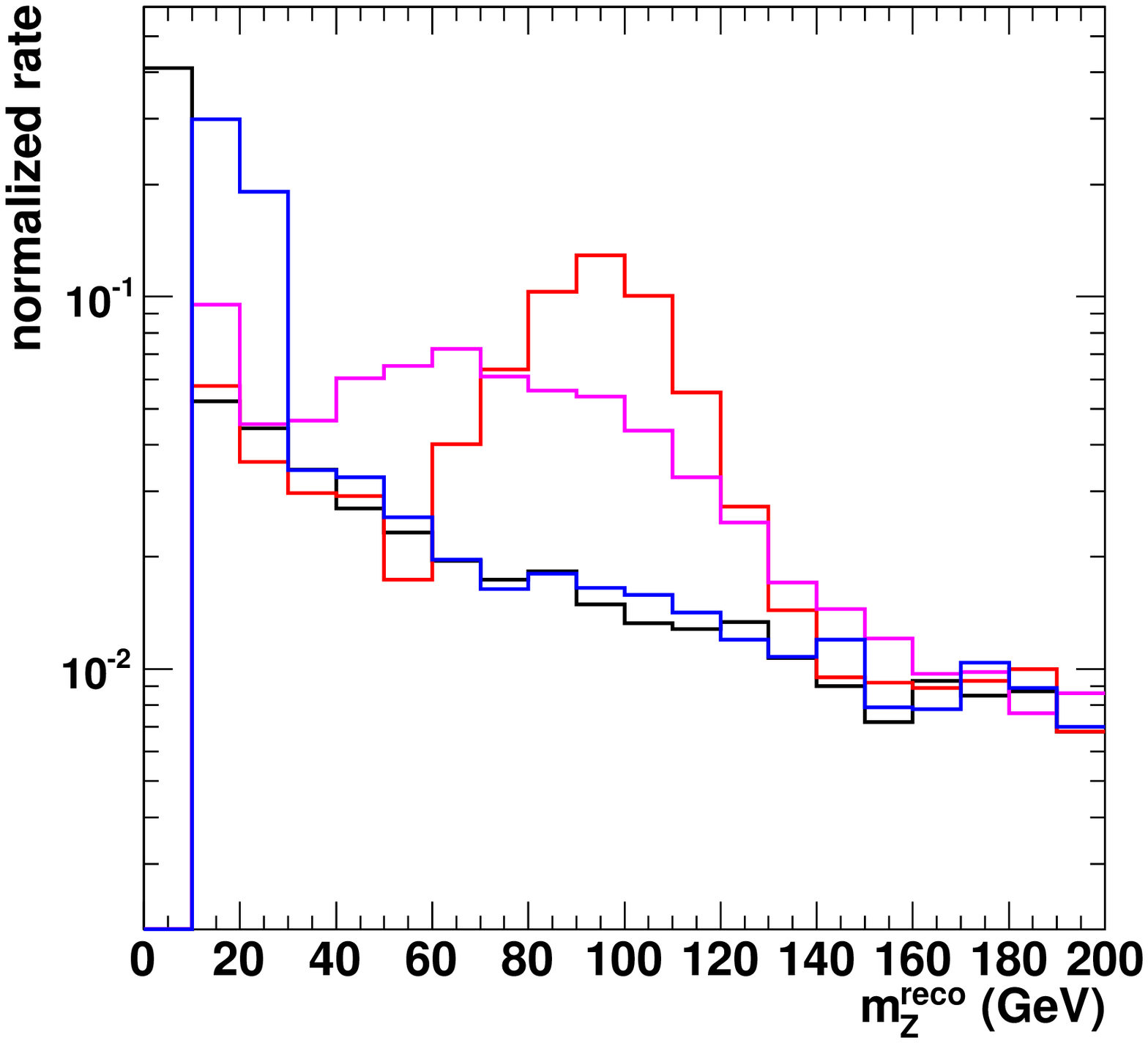}
\caption{\it Distributions of the reconstructed hadronic $Z$ mass for 3 TeV $Z' \to Zh \to (q\bar q)(l\nu l\bar\nu)$ (left) and quark-jets with $p_T \simeq 1500$ GeV (right) processed through the BDRS mass-drop procedure.  Displayed are particle-level (black), simple ECAL and HCAL cells (red), ECAL cells rescaled per containing HCAL cell (pink), and ECAL cells rescaled per containing HCAL cluster (blue).}
\label{fig:ZjetCAL}
\end{center}
\end{figure}

\begin{figure}[tp]
\begin{center}
\epsfxsize=0.44\textwidth\epsfbox{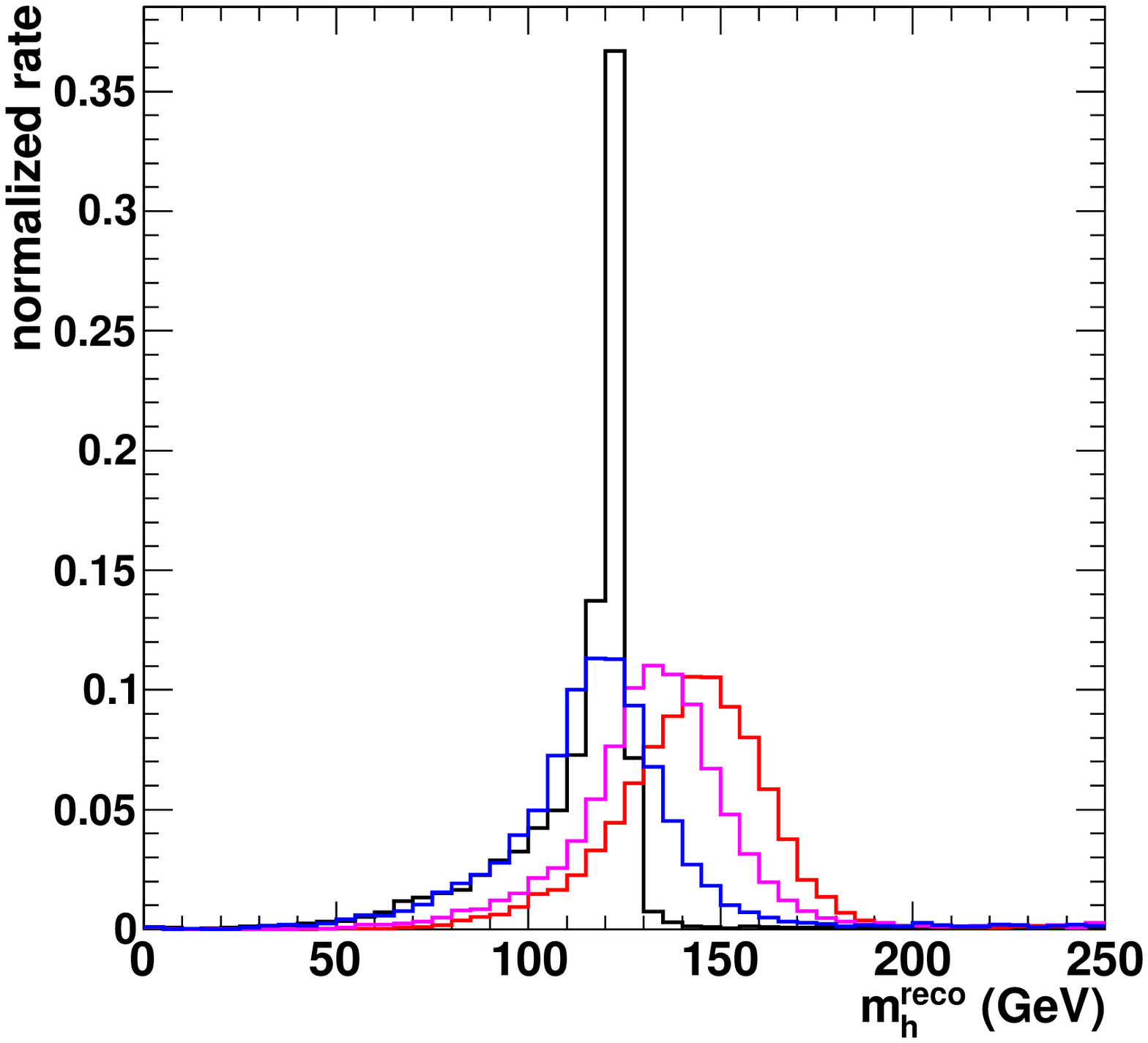}
\epsfxsize=0.44\textwidth\epsfbox{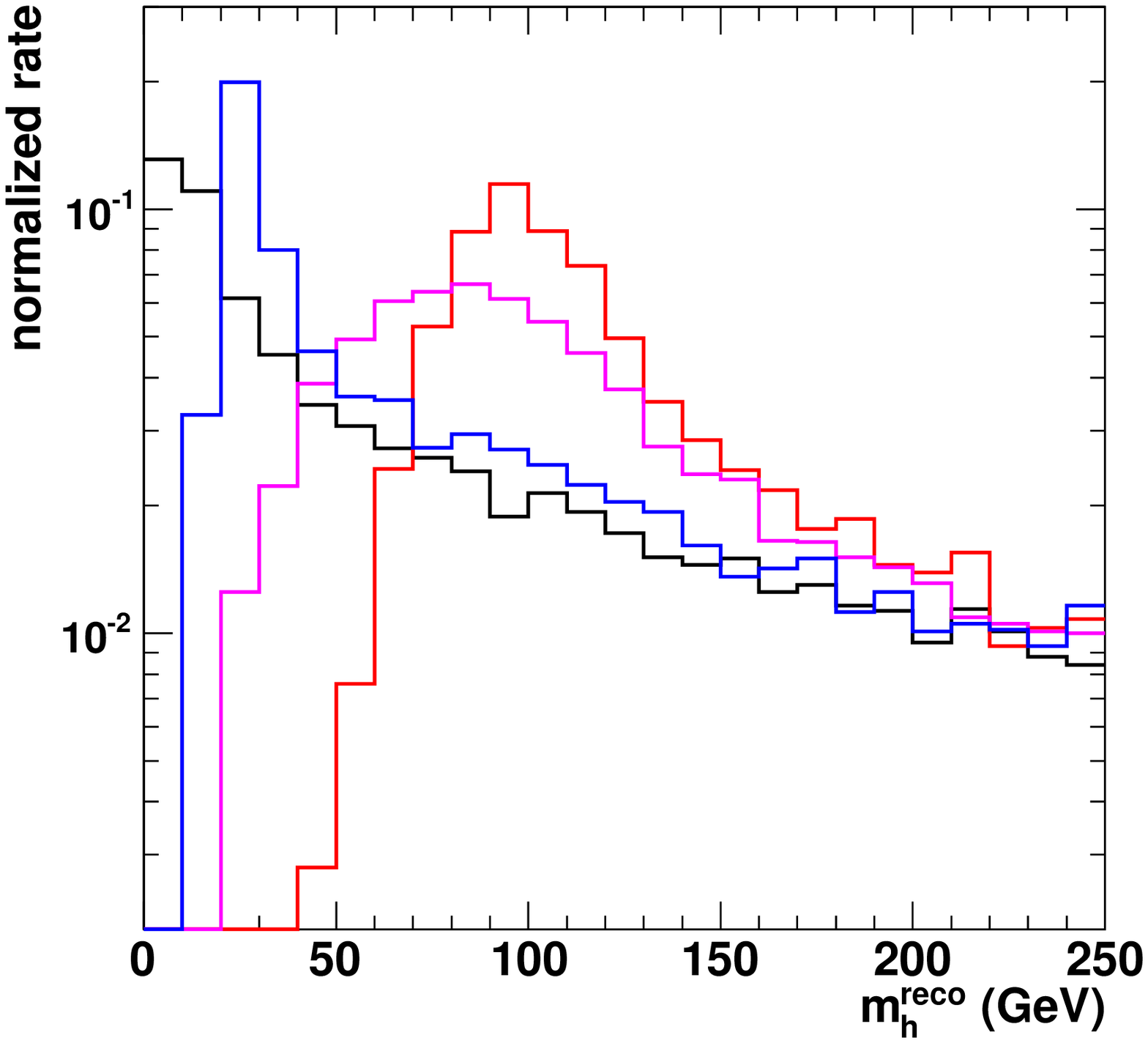}
\caption{\it Distributions of the reconstructed fully hadronic 125 GeV Higgs mass for 3 TeV $Z' \to Zh \to (\nu\bar\nu)(q\bar{q} 'q\bar{q}')$ (left) and quark-jets with $p_T \simeq 1500$ GeV (right) processed through our 4-body declustering procedure.  Displayed are particle-level (black), simple ECAL and HCAL cells (red), ECAL cells rescaled per containing HCAL cell (pink), and ECAL cells rescaled per containing HCAL cluster (blue).}
\label{fig:HjetCAL}
\end{center}
\end{figure}

In a realistic detector, substructure-sensitive observables are affected by finite energy resolution and finite spatial resolution.  Because the jets and subjets that we consider are very energetic, and calorimeter energy measurements become better with increasing energy (down to few-percent resolution), we do not expect that energy resolution effects will be a major obstacle.  Spatial resolution, on the other hand, may pose a very serious fundamental limitation.  For example, in our most energetic $Z$-jets and diboson-jets, individual subjets may lie inside of a single hadronic calorimeter cell.

The ATLAS and CMS detectors consist of multiple subdetectors, of which the HCAL is the spatially coarsest.  By combining tracker, ECAL, and HCAL information together, it is possible to extract a very detailed picture of the energy flow of the event~\cite{CMSpf}.  However, it is known that tracking becomes less reliable as jet energies increase, due to the increasing density of hits and the decreasing track curvatures.  This may not be a fatal issue, but to be conservative we consider only the ECAL and HCAL.

In a previous paper~\cite{Katz:2010mr}, we suggested a simple method to hybridize the ECAL's 4--5 times better spatial resolution with the full ECAL+HCAL energy measurement:  Within each HCAL cell and its associated block of ECAL cells, rescale the ECAL cell energies so that the sum of ECAL cells matches the full ECAL+HCAL energy.  These rescaled ECAL cells then serve as the 4-vectors input into jet clustering and jet substructure.  We found that this technique was very effective at correcting off the mass-distorting effects of the HCAL geometry, at least in the context of a toy calorimeter model with perfect energy-sampling cells.

Here, we add some important extra layers of reality to our toy calorimeter by approximately accounting for two additional effects:  1) The energy deposited by each particle will shower into several cells, and 2) Hadrons will often deposit energy in the ECAL due to nuclear interactions.  The two effects partially cancel each other, as the former introduces spurious substructure, whereas the latter increases the share of jet energy collected in the more finely-segmented ECAL.

To implement the spatial energy smearing, we continuously distribute the energy of each particle using the profile parametrization in~\cite{Nakamura:2010zzi}, setting the Moli{\`e}re radius equal to one cell unit (individually for ECAL and HCAL).  (See~\cite{LochModel} for a similar model.)  For simplicity, we apply the smearing in $\eta$--$\phi$ space rather than in real space, which should be an adequate approximation except at high $\eta$.  ECAL and HCAL cells are respectively set to 0.02 and 0.1 unit across.  Our spatial smearing is specifically designed to furnish a good model of the CMS ECAL, and is likely somewhat pessimistic for the HCAL of either experiment.

To implement the hadronic ECAL deposits, we apply a 65\% probability for charged hadrons above an energy of 5~GeV to deposit some energy.  The fraction of energy is taken from a simple linearly-falling distribution that hits zero probability at 100\% deposition.  These provide a coarse model of the CMS ECAL response to charged hadrons as observed in test beam data~\cite{CMSecal,CMScalo}.  Including this effect increases the typical EM fraction of a jet from about 25\% (mainly $\pi^0\to\gamma\gamma$) to 40\%.  In either case, the RMS of the distribution is comparable to half of the mean.  More realistic jet EM fractions in the complete detector would be 55\% for CMS~\cite{CMSemfrac} and 80\% for ATLAS~\cite{ATLASemfrac}, with similar relative RMS.

We can see the effects of our calorimeter model on $p_T \simeq 1500$~GeV $Z$-jet and Higgs-jet mass reconstructions in Figs.~\ref{fig:ZjetCAL} and~\ref{fig:HjetCAL}, respectively.  (Examples of diboson-jet event displays appear in Fig.~\ref{fig:legos}.)  If we simply add up the ECAL and HCAL cell 4-vectors, there is an obvious degradation of the signal peaks.  Perhaps more worrisome, the $p_T$ scale and spatial resolution scale interplay to create an artificial mass feature in the background QCD jets at about 100~GeV.  Applying our naive ECAL rescaling brings us closer to particle-level, but there is still an obvious bias.  To achieve a better correction, we can apply a slight modification to the ECAL rescaling.  We first ``undo'' the spatial smearing effects by clustering HCAL cells with the anti-$k_T$ algorithm~\cite{Cacciari:2008gp} with $R = 0.17$ (smaller than 2 cells but bigger than $\sqrt{2}$ cells).  We then rescale all ECAL cells associated with each {\it cluster} of HCAL cells, rather than cell-by-cell.  The result of this procedure also appears in Figs.~\ref{fig:ZjetCAL} and~\ref{fig:HjetCAL}, and clearly indicates that most of the remaining bias has been removed.  Not only is the signal peak approximately restored, but the artificial intrinsic mass scale of the background jets has been moved down to $O$(20~GeV).

The quality of the final mass reconstruction may seem somewhat surprising, given that the fraction of energy deposited in the ECAL fluctuates a lot.  The mass of a quasi-collinear subjet pair goes like $p_T\Delta R\sqrt{z(1-z)}$, where $p_T$ is the total transverse momentum of the pair, $\Delta R$ is their $\eta$--$\phi$ distance, and $z = [0,0.5]$ parametrizes their energy sharing.  With our procedure, $p_T$ and $\Delta R$ are very well-measured, and the potential problem is in the $z$ measurement.  For subjets with equal energy sharing, the $\sqrt{z(1-z)}$ factor is 0.5.  Even if we mismeasured the energy sharing by a factor of 2, so that $z\to0.25$, the result shifts by only 14\%.  More asymmetric splittings suffer more:  $z = 0.1$ mismeasured as $z = 0.05$ changes the mass by 27\%.  However, averaging over many subjet configurations and EM fractions, the net mass smearing effect is rather modest.

There are many other aspects of the real calorimeters that we did not consider, such as shower fluctuations, noise, showering in the tracker, and magnetic field effects.  We do not expect that these will significantly alter our conclusions, but they would certainly need to be accounted for to verify that our cluster-rescaling trick works as advertised.  We also note that this type of procedure should be useful for other jet substructure applications that require fine angular resolution, such as high-$p_T$ top-tagging.



\bibliography{lit}
\bibliographystyle{apsper}

\end{document}